\documentclass[aps,pre,citeautoscript,twocolumn,superscriptaddress,nofootinbib,nobibnotes,10pt]{revtex4-1}
\usepackage{graphicx}
\usepackage{dcolumn}
\usepackage{bm}
\usepackage{amsmath,amssymb}
\usepackage[utf8]{inputenc}
\usepackage[T1]{fontenc}
\usepackage{mathptmx}
\usepackage{etoolbox}
\usepackage{lineno}
\makeatletter
\def\@email#1#2{%
 \endgroup
 \patchcmd{\titleblock@produce}
  {\frontmatter@RRAPformat}
  {\frontmatter@RRAPformat{\produce@RRAP{*#1\href{mailto:#2}{#2}}}\frontmatter@RRAPformat}
  {}{}
}%
\makeatother

\begin{document}
\preprint{AIP/123-QED}

\title{Impact of phase modulation on the dynamics of temporal localized structures in injected Kerr microcavities}
\author{Marc Hunkemöller}
\affiliation{Institute of Climate and Energy Systems, Forschungszentrum Jülich, Wilhelm-Johnen-Str., 52428 Jülich, Germany}
\affiliation{Institute for Theoretical Physics, University of Münster, Wilhelm-Klemm-Str. 9, 48149 Münster,
Germany}%
\author{Thomas G. Seidel}%
\affiliation{Institute for Theoretical Physics, University of Münster, Wilhelm-Klemm-Str. 9, 48149 Münster, Germany}%

\author{Julien Javaloyes}
\affiliation{Departament de Física, Universitat de les Illes Balears, \& Institute of Applied Computing and Community Code (IAC-3),
C/ Valldemossa km 7.5, 07122 Mallorca, Spain
}%

\author{Svetlana V. Gurevich}
\affiliation{Institute for Theoretical Physics, University of Münster, Wilhelm-Klemm-Str. 9, 48149 Münster,
Germany}%
\affiliation{
Center for Data Science and Complexity (CDSC), University of Münster, Corrensstr. 2, 48149 Münster, Germany}

\date{\today}

\begin{abstract}
We theoretically investigate how phase modulation alters the dynamics of temporal localized structures (TLSs) in vertically emitting Kerr micro-cavities under detuned optical injection operating in the normal dispersion regime. We show that the emergence of TLSs in general is governed by a synchronization between the imposed modulation and the intrinsic pulse dynamics. We perform a multi-parameter bifurcation analysis of the underlying delay–algebraic equation model in the uniform field limit and demonstrate that weakly nonlinear and dissipative Hermite--Gauss modes shape the dynamics of dark TLSs, leading to a complex hybrid bifurcation structure. Beyond the uniform field limit, both bright and dark modulated TLSs are shown to exist and to occupy distinct equilibrium positions within the cavity. An effective equation of motion for the TLS positions is derived, showing a good agreement with the full model.
\end{abstract}

\maketitle

\begin{quotation}
We study the impact of phase modulation on the nonlinear dynamics of temporal localized structures in optically injected Kerr micro-cavities operating in the normal dispersion regime. The formation of localized states is shown to arise from a synchronization between the external modulation and the intrinsic pulse dynamics, leading to modulation-locked solutions over a finite parameter range. A multi-parameter bifurcation analysis reveals a complex hybrid bifurcation structure in which weakly nonlinear dissipative Hermite--Gauss modes govern the dynamics of dark localized states. Away from the uniform field limit, both bright and dark modulated localized structures coexist and occupy distinct equilibrium positions, whose dynamics are well described by an effective equation of motion.
\end{quotation}

\section{\label{sec:intro} Introduction}

Optical frequency combs (OFCs) are highly coherent light sources with spectra consisting of a series of discrete and evenly spaced spectral lines. Their discovery have had a significat impact on a variety of fields including metrology, high-precision optical spectroscopy and optical communications, see~\cite{D-JOSAB-10,PPR-PR-18} for reviews. Prominent approaches of OFCs generation leverage the temporal cavity solitons obtained in mode-locked vertical external-cavity surface-emitting lasers~\cite{tropper04,LMB-OE-10} and injected passive Kerr resonators such as fiber loops~\cite{LCK-NAP-10} and microrings~\cite{HBJ-NAP-14}. The physical processes of the OFC formation in Kerr resonators depend on the respective signs of the dispersion and Kerr nonlinearity and, in the uniform field limit (UFL), where the intracavity power is relatively low and the overall cavity detuning remains small during propagation, the mean-field approximation results in the Lugiato-Lefever Equation (LLE) or coupled mode models~\cite{LL-PRL-87,HTW-OC-92,PPR-PR-18,CSY-PRL-10,CY-PRA-10,CM-PRA-13,CRSE-OL-13,HBJ-NAP-14,CPP-PRL-21}. Here, both bright and dark temporal localized states (TLSs) can be observed in the anomalous and normal dispersion regimes, respectively~\cite{LCK-NAP-10,LLK_OE_15,XXL_NP_15}. In particular, in the normal dispersion regime, they can be build up from domain walls that connect domains of the corresponding lower and higher continuous wave (CW) background intensities~\cite{PKG_PRA_16,GWM_EPJD_17} and their coexistence is ensured by e.g., the presence of third-order dispersion (TOD)~\cite{PGL-OL-14,PGG_PRA_17} or other high order effects~\cite{TG-OL-10}.

Recently, an alternative method for the generation of phase-locked tunable OFCs in
optically injected vertical-emission Kerr–Gires–Tournois interferometers (KGTI) has been proposed using a model relying on delay algebraic equations (DAE)~\cite{SPV-OL-19,SJG-OL-22,KSG-OL-22,KSJ-CHA-23,KGJ-CSF-26}. In the normal dispersion regime, the DAE model successfully predicts the formation of both bright and dark TLSs formed via the locking of domain walls connecting the high and low intensity levels of the injected micro-cavity. Further, it was shown that in the regime of small cavity losses and weak injection, the dynamics of the KGTI system can be described by the LLE with third-order dispersion~\cite{SGJ-PRL-22}. Finally, operating the KGTI system far beyond the mean field limit, the existence of a new type of short high-intensity TLSs was recently reported~\cite{SJG-OL-24}.

Various modulation techniques have been suggested theoretically and applied experimentally to gain control over the TLSs dynamics in Kerr resonators. In particular, inhomogeneity in pumping caused by an intensity or a phase modulation of the driving field was successfully employed over the last decade~\cite{OTC-PRA-14,JEC-NAC-15,CSE-Optica-18,Hendry2018,TFHP-PRA-19,DHT-OL-24}, resulting in e.g., the control on the TLSs position~\cite{JEC-NAC-15}, enhancement of the pump-to-soliton conversion efficiency~\cite{OLH-NatPhot-17}, or formation of hybrid TLSs in non-Hermitian Kerr cavities~\cite{IMB-PRL-24}. Alternatively, the intracavity field can be directly modulated using e.g., an electro-optical modulator, see, e.g.~\cite{BOP-PRL-12,TTK-PRA-20}. The associated parabolic potential was shown to stabilize TLSs in anomalous dispersion regime~\cite{SPRF-OL-22}, induce high-order TLSs in the normal dispersion regime~\cite{SWPR-OL-23}, and leads to the formation of high-order breathers and chaoticons~\cite{SPRF-CSF-23}.

In this paper we theoretically investigate the impact of the phase modulation onto the dynamics of TLSs found in the KGTI system operated in the normal dispersion regime. The phase modulation is introduced by periodically moving the external feedback mirror which causes a small variation in the length of the external cavity. In particular, we show that the formation of TLSs in the modulated KGTI system can be interpreted as a synchronization process between the external phase modulation and the intrinsic pulse dynamics and demonstrate the presence of a synchronization region, in which TLSs are locked to the external modulation. Next we focus on the dynamics of the system in question in the UFL, where the dynamics is governed by the Lugiato-Lefever equation with TOD. We perform a multi-parameter bifurcation study of the full DAE model in the absence of TOD and show, how weakly nonlinear and dissipative Hermite--Gauss (HG) modes enslave the dynamics of dark TLSs leading to a complex hybrid bifurcation structure. Further, in the parameter regimes far from the UFL, we demonstrate that the modulated bright and dark TLSs can be multistable  and occupy distinct positions in the cavity relative to the phase modulation profile. The multistability again can be understood as a synchronization process, where several locking regions overlap for a certain parameter regions. To elucidate the mechanisms behind the position selection, we derived an effective equation of motion for the respective TLSs positions, showing a good agreement with the results obtained from the full DAE model.

\section{\label{sec:model} Model Equations}
\begin{figure}
\includegraphics{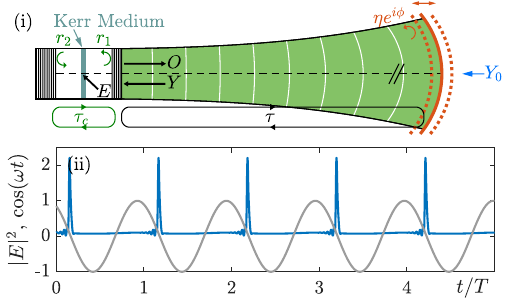}
\caption{\label{fig:1}(i) Schematic of a micro-cavity containing a Kerr medium. The micro-cavity has a round-trip time $\tau_\mathrm{c}$ and is coupled to an external cavity of the round-trip time $\tau$, closed by a mirror (orange) with reflectivity $\eta$ and phase shift $\phi$. The system is driven by a monomode injection field $Y_0$, and the feedback phase is modulated via periodic mirror motion.
(ii) Intensity of a single temporal pulse circulating in the external cavity (blue) obtained by integrating Eqs.~\eqref{subeq:KGTI_E} and \eqref{subeq:KGTI_Y} with $m=0.12$ and $\omega=0.0619$ plotted together with the feedback phase modulation (gray). Other parameters are $(\delta,\,h,\,\eta\,,\varphi_0\,,Y_0\,,\tau)=(1.5\,,2\,,0.75\,,0\,,0.55\,,100)$.}
\end{figure}

A schematic of the KGTI system in question is depicted in Fig.~\ref{fig:1}~(i). It is composed of a monomode microdisk cavity of a few micrometers in length and with round-trip time $\tau_c$ containing a Kerr medium and bounded by two distributed Bragg mirrors with reflectivities $r_{1,2}$. The micro-cavity is coupled to a long external cavity of a few centimeters with round-trip time $\tau \gg \tau_\mathrm{c}$ which is closed by a mirror with reflectivity $\eta$. The coupling efficiency between the two cavities is determined by the factor $h=h(r_1,\,r_2) = (1+|r_2|)(1-|r_1|)/(1-|r_1||r_2|)$. Here, we consider the case of a perfectly reflecting bottom mirror, i.e., $h(r_1, 1) = 2$, the so-called Gires-Tournois interferometer regime~\cite{GT-CRA-64}. The system is driven by a monomode injection beam with amplitude $Y_0$ and frequency $\omega_{Y_0}$. The injection is detuned from the micro-cavity resonance $\omega_\mathrm{mc}$ resulting in the detuning $\delta=\omega_\mathrm{mc}-\omega_{Y_0}$. Further, we define $E$ and $Y$ the slowly varying envelope of electric fields in the micro-cavity and the external cavity, respectively. The field cavity enhancement can be conveniently scaled out using the Stokes relations allowing $E$ and $Y$ to be of the same order of magnitude~\cite{SPV-OL-19,KGJ-CSF-26}. This leads to a simple input-output relation $O = E - Y$.
%
%
The output $O$ is re-injected into the external cavity after one roundtrip with an attenuation factor $\eta e^{i\varphi}$, cf. Refs.~\onlinecite{SPV-OL-19,SJG-OL-22,KGJ-CSF-26}. Here, the total phase $\varphi$ is composed of the phase accumulated during propagation, $\omega_{Y_0} \tau$, and the phase shift introduced by the feedback mirror, $\phi$, i.e., $\varphi=\phi+\omega_{Y_0}\tau$.
%
%

\begin{figure*}
\includegraphics{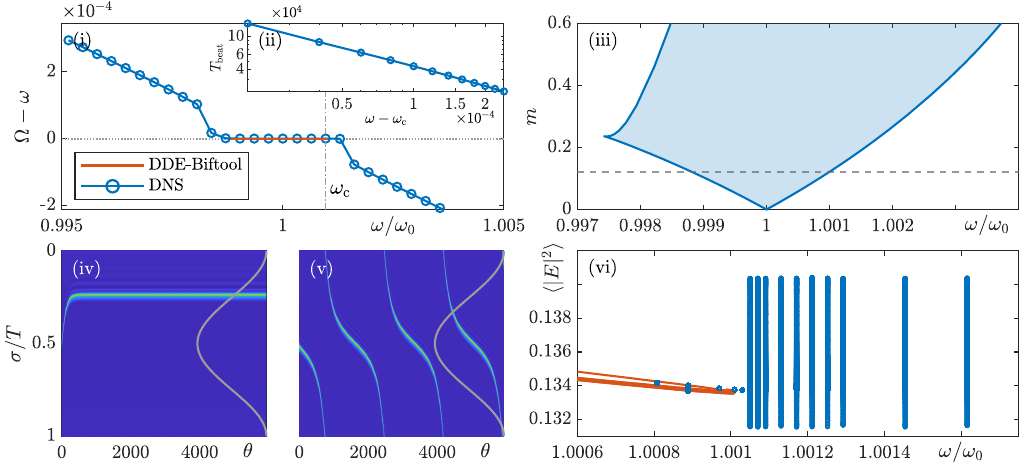}
\caption{\label{fig:2}(i) Difference of the the repetition frequency $\Omega$ of the pulses in the KGTI system~\eqref{eq:whole_KGTI} and the frequency of the phase modulation $\omega$. The data are obtained numerically (blue circles) and by path continuation (orange) for $m=0.12$. The horizontal axis is scaled by the natural frequency of the pulses $\omega_0$. (ii) The scaling of the beat period $T_\mathrm{beat}$ of the quasi-periodic dynamics near the critical frequency $\omega_\mathrm{c}$ in a double logarithmic scale. (iii) The Arnold tongue indicating the synchronization region as a function of the amplitude $m$ and frequency $\omega$ of the phase modulation. The dashed line corresponds to the panel (i). (iv), (v) present synchronized ($\omega=0.0619$) and quasi-periodic ($\omega=0.06198$) dynamics, respectively, using a two-time representation. The gray line indicates the phase modulation. (vi) The orange branch represents the path continuation of a synchronized solution (stable/unstable solutions in bold/thin), while the blue dots correspond to numerical time simulations, showing the integrated intensity in each round-trip. Other parameters are as in Fig.~\ref{fig:1}.}
\end{figure*}

Finally, we introduce phase modulation to the KGTI system by periodically moving the external feedback mirror which causes a small variation in the length of the external cavity, i.e. $\tau(t) = \tau + \Delta \tau(t)$. Here, $\Delta \tau(t)$ is small modulation such that $\Delta \tau(t)/\tau\ll1$. Consequently, this alters the propagation phase per round-trip such that the total phase reads $\varphi(t) =  \phi + \omega_{Y_0}\tau(t)$. A natural choice for the modulation $\Delta \tau(t)$ is for it to be harmonic which can be achieved by, e.g., using a piezoelectric motor. With a modulation amplitude $m$ and frequency $\omega$, the total modulated phase reads
\begin{equation}
    \varphi(t)=\varphi_0+m\cos(\omega\,t),
    \label{eq:feedback_phase}
\end{equation}
where $\varphi_0 = \phi + \omega_{Y_0} \tau$ is the non-modulated feedback phase and $m=\omega_{Y_0}\Delta\tau$.
%
Following the methods developed in~\cite{SPV-OL-19,SJG-OL-22,KSG-OL-22,KSJ-CHA-23,SJG-OL-24,KGJ-CSF-26}, we can write the evolution of the normalized slowly varying field envelopes $E$ and $Y$ as follows
\begin{subequations}
\label{eq:whole_KGTI}
\begin{eqnarray}
\dot{E}&=&\left[-1+i\left(\left|E\right|^2-\delta\right)\right]E+h\,Y\,,\label{subeq:KGTI_E}\\
Y&=&\eta e^{i\varphi(t)}\left[E(t-\tau)-Y(t-\tau)\right]+\sqrt{1-\eta^2}\, Y_{0},\label{subeq:KGTI_Y}
\end{eqnarray}
\end{subequations}
where the feedback phase $\varphi(t)$ is given by Eq.~\eqref{eq:feedback_phase} and $h=h(r_1, 1)=2$. The effects of the modulation on the delayed output can be neglected as $O\left(t-\tau\left(t\right)\right)\simeq O\left(t-\tau\right)+\mathcal{O}\left(\Delta\tau\left(t\right)/\tau\right)$.

The coupling between the intra- and external cavity fields is given by the DAE~\eqref{subeq:KGTI_Y} which takes into account all the multiple reflections in a possibly high finesse external cavity. Hence, both group delay dispersion and third-order dispersion (TOD) are naturally captured by the system~\eqref{eq:whole_KGTI}~\cite{SPV-OL-19,SCM-PRL-19}, see also~Ref.~\onlinecite{VD-PRE-24}. Note that the chosen modeling approach is similar to the one used for microdisk mode-locked lasers; see, e.g. Refs.~\onlinecite{MB-JQE-05,CSV-OL-18,SHJ-PRAp-20,HGJ-OL-21}.
%
%

The amount and sign of the second order dispersion, which typically constitute the dominant effect away from resonance, can be tuned by selecting the operating frequency relative to the cavity resonance. In contrast, in the vicinity of the resonance the second-order dispersion vanishes and changes sign, causing TOD to become the leading contribution. Due to TOD, the resulting time-periodic solutions of the system~\eqref{eq:whole_KGTI} are in general asymmetrical and can possess strong oscillatory tails.
In particular, in the absence of the phase modulation and in the normal dispersion regime, the bistable response resulting from a Kerr medium allows to find bright and dark TLSs that are formed via the locking of domain walls connecting the high-and low-intensity levels of the injected micro-cavity, leading to
complex shape multistability, cf. Refs.~\onlinecite{SPV-OL-19,SJG-OL-22,KSG-OL-22,KSJ-CHA-23,SJG-OL-24,KGJ-CSF-26}. Figure \ref{fig:1}~(ii) shows the exemplary dynamics over several round-trips of a single temporal pulse per cycle (blue) and the corresponding modulated feedback phase (gray) obtained by numerical integration of the system~\eqref{eq:whole_KGTI} for $m\neq0$. Note that TLSs are located not at the minima of the corresponding periodic potential landscape caused by the phase modulation. In what follows we analyze the impact of the phase modulation on the dynamics and stability of TLSs of the DAE~\eqref{eq:whole_KGTI} in the normal dispersion regime.
%
%

\section{\label{sec:synchronization} Synchronization by external Force}


The appearance of a single temporal pulse per round-trip presented in Fig.~\ref{fig:1}~(ii) can be understood as a synchronization process. Indeed, the phase-modulated KGTI system~\eqref{eq:whole_KGTI} possesses two  frequencies. The first is the external modulation frequency $\omega$, while the second is the repetition rate $\Omega$ of the driven corresponding periodic solution. However, it can lock~\cite{synchrobook} to the external frequency such that there is e.g., exactly one pulse per modulation period $T = 2\pi/\omega$, and, after a transient, the resulting pulse occurs consistently at the same point within each period. This resulting synchronized behavior is shown in Fig.~\ref{fig:2}~(iv), where the color map represents the intensity of the $E$-field. Here, a full temporal evolution of the one-pulse solution shown in Fig.~\ref{fig:1}~(ii) is presented using a two-time representation~\cite{AGL-PRA-92,GP-PRL-96,FG-PRE-20}, where the fast time scale $\sigma$ governs the dynamics within one round-trip, whereas the slow scale $\theta$ describes the dynamics from one round-trip to the next, i.e., $t = \sigma + \theta T$, $\sigma\in[0,\,T)$. Again, the gray line indicates the phase modulation profile.


In order to further quantify the occurring synchronization process we examine numerically the detuning between the pulse repetition frequency $\Omega$ and the phase modulation frequency $\omega$, scaled by the natural frequency of the pulses $\omega_0$, cf. blue circles in Fig.~\ref{fig:2}~(i). As expected~\cite{synchrobook}, the detuning $\Omega-\omega$ vanishes when synchronization occurs. The resulting synchronized solution are periodic orbits of the system~\eqref{eq:whole_KGTI} and can be analyzed with the path-continuation methods within the recently developed extension of DDE-BIFTOOL~\cite{DDEBT} that allows for the bifurcation analysis of algebraic and neutral delayed equations. The result is depicted in Fig.~\ref{fig:2}~(i) in orange. One can see that the branch of synchronized solutions forms a very narrow loop, bounded by fold bifurcations at both ends. The loop structure becomes more visible by using another measure such as the integrated intensity, cf. Fig.~\ref{fig:2}~(vi). Here only a part of the branch is presented and bold (thin) orange lines indicate a stable (unstable) solutions, respectively. An example of the stable solution is depicted in Fig.~\ref{fig:2}~(iv), whereas for unstable solutions, the pulse is shifted by half a period in the $\omega$-direction, positioned on the rising edge of the modulation.
%
%

The synchronization region shown in Fig.~\ref{fig:2}~(i) can be extended to a second dimension in the modulation amplitude $m$ forming a characteristic Arnold tongue~\cite{synchrobook}, see Fig.~\ref{fig:2}~(iii). Here, the dashed horizontal line represents the cut presented in Fig.~\ref{fig:2}~(i). The boundaries of the Arnold tongue are determined continuing the fold bifurcations that define the edges of the synchronized solution branch (cf. Fig.~\ref{fig:2}~(vi)) in $m$. One can see that for small modulation amplitudes, the boundaries of the Arnold tongue exhibit a nearly linear dependence but for large $m$ values the triangular shape of the tongue is modified, indicating that entrainment of the pulses is easier at frequencies higher than their natural frequency. Note that the observed frequency locking is not limited to the case of 1:1 synchronization and, in general, high order $n:m$ tongues, where the condition $n\,\omega=m\,\Omega$ is fulfilled, can be found.

In contrast to the periodic synchronized state, in an unsynchronized case, the pulses are quasi-periodic orbits. Here, the modulation frequency $\omega$ deviates too much from the natural solution frequency $\omega_0$ or the modulation amplitude $m$ is insufficient to entrain the pulse to the periodic modulation. Nevertheless, the pulses still respond to the modulation, leading to non-uniform motion. Depending on the exact position within the modulation, the pulses are either accelerated or slowed down. In the two-time representation, as shown in Fig.~\ref{fig:2}~(v), the modulation causes the pulse to propagate periodically through the $\sigma$-direction with a varying speed. This motion is again periodic with period $T_\mathrm{beat}$ , which allows us to identify a beat frequency $\Omega_\mathrm{beat}=2\pi/T_\mathrm{beat}$. The interplay between the two frequencies $\omega$ and $\Omega_\mathrm{beat}$ results in the quasi-periodic motion.


For the quasi-periodic regime, only numerical time simulations can be employed, shown by the blue dots in Fig.~\ref{fig:2}~(i) and (vi). In particular, panel (vi) indicates that the integrated intensity varies significantly between the round-trips, and in panel (i) we can observe that the detuning $\Omega-\omega$ quickly increases above the critical frequency $\omega_c$ where the fold bifurcations of the branch of synchronized solutions (orange) occur. Analyzing the period of the beating in detail (cf.~Fig.~\ref{fig:2}~(ii)) reveals that $T_\mathrm{beat}$ diverges close to $\omega_{c}$, and, in particular, exhibits an $\left(\omega-\omega_c\right)^{-1/2}$-dependence. This characteristic scaling law reveals that the bifurcation indicating the transition between synchronized and unsynchronized solutions is of a saddle-node infinite period (SNIPER) type.

\section{\label{sec:ufl} Uniform-field limit}
\begin{figure*}
\includegraphics{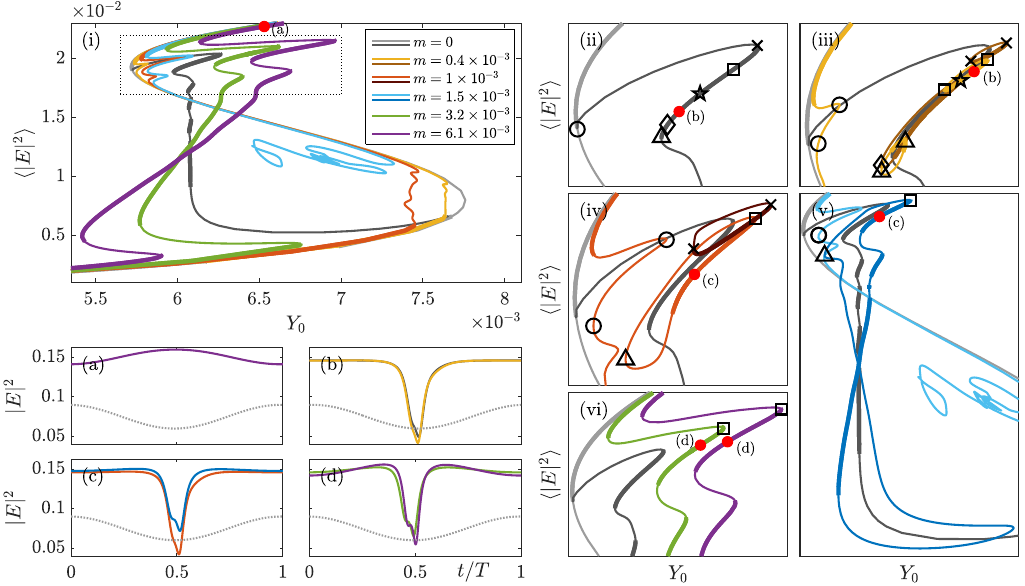}
\caption{\label{fig:3}
Bifurcation diagrams of Eqs.~\eqref{subeq:KGTI_E} and \eqref{subeq:KGTI_Y} showing the integrated intensity $\langle|E|^2\rangle$ as a function of the injection $Y_0$ for different amplitudes of the modulation $m$. Panel (i) shows the CW (light gray) and the TLS (dark gray) branches in the background for $m=0$. Additionally, the modulated CW branches for $m=\{0.4,\,1,\,1.5,\,3.2,\,6.1\}\times10^{-3}$ are depicted in yellow, orange, blue, green and purple, respectively. The region marked with a dotted rectangle is shown in detail, together with the corresponding TLS branches in panels (ii) - (vi). Different black markers indicate points traced for increasing $m$. Panels (a)-(d) present exemplary profiles corresponding to the points marked with red dots in panels (i)-(vi), whereas the dotted gray line indicates the phase modulation. The system parameters are $(\delta,h,\eta,\varphi_0,\tau,\omega)=\left(1/\sqrt{3},2,0.99,0.3235\pi,300,0.0208\right)$.
}
\end{figure*}

\subsection{Normal form partial differential equation}
Following our analysis in the framework of synchronization, we proceed  performing a detailed bifurcation study of TLSs found in Eqs.~\eqref{eq:whole_KGTI} for $m\neq0$ in the normal dispersion regime, i.e., for $\delta>0$. We start our analysis from a so-called uniform field limit (UFL), characterized by small cavity losses, small phase shifts and weak injection. In this regime, the dynamics of the KGTI system~\eqref{eq:whole_KGTI} in the long cavity limit can be well-described by the Lugiato-Levefer equation~\cite{LL-PRL-87,HTW-OC-92} with third-order dispersion~\cite{SGJ-PRL-22}
\begin{equation}
    \left(i\partial_\theta+\tilde{\varphi}(\sigma)-\frac{\beta_2}{2}\partial_\sigma^2-i\frac{\beta_3}{6}\partial_\sigma^3+\gamma|E|^2\right)E=F(E),
    \label{eq:LLE}
\end{equation}
where again $\sigma$ and $\theta$ correspond to fast and slow time scales, respectively, $F(E)=i(\eta-1)E+iY_0\sqrt{8(1-\eta)}/(1+i\delta)$ contains both losses and injection, $\beta_2=4\delta/(1+\delta^2)^2$, $\beta_3=4(3\delta^2-1)/(1+\delta^2)^3$ and $\gamma=2/(1+\delta^2)$. The effective round-trip phase  $\tilde{\varphi}(\sigma)=\varphi(\sigma)-2\arctan(\delta)$ is now $\sigma$-dependent and corresponds to the weak ''spatial`` potential.

We note that the second- and third-order dispersion coefficients $\beta_2$ and $\beta_3$ are controlled by the detuning $\delta$ and, for the sake of simplicity, we choose first $\delta=1/\sqrt{3}$, where $\beta_3$ vanishes.
Hence, without modulation of the feedback phase, i.e., for $m=0$, the solutions of the DAE system~\eqref{eq:whole_KGTI} in the UFL are either continuous wave (CW) solutions, arranged on a S-shaped branch exhibiting bistability of a high- and low-intensity solution, or TLSs which form a collapsed snaking branch~\cite{SJG-OL-22} connecting the two fold bifurcation points of the CW solution, see light and dark gray lines in the bifurcation diagram in the injection $Y_0$ presented in Fig.~\ref{fig:3}~(i) for CW and TLSs solution branches, respectively. Here, solid (thin) lines correspond to stable (unstable) TLS solutions, respectively. Note that since $\beta_3=0$, only dark TLSs are present~\cite{PGL-OL-14}.

\subsection{Bifurcation analysis in the UFL limit}
\label{sec:bif_ufl}
We introduce the phase modulation and gradually increase its amplitude from $m=0.4\times10^{-3}$ to $6.1\times10^{-3}$, while keeping the modulation frequency fixed at $\omega=0.0208$. As $m$ increases, the initially S-shaped CW branch (light gray) becomes increasingly deformed at the fold bifurcation points, see Fig.~\ref{fig:3}~(i). At the upper fold, the resulting shape begins to resemble the collapsed snaking structure characteristic of the TLS branch (cf. also panels (ii)-(iv) for small $m$ values), whereas at the lower fold, the deformation is irregular. At $m=1.5\times10^{-3}$, the modulated CW branch (light blue) reconnects to some unstable branch (not shown). Eventually, at $m\geq3.2\times10^{-3}$, the modulated CW branch (green, purple) reconnects back to the stable low intensity solution, resulting in an overall shape that resembles a tilted version of the collapsed snaking seen for the TLS branch at $m=0$, cf. panel (vi). Note that for $m=0$, the profile of the CW solution is flat, but as $m$ increases, the intensity becomes modulated, see Fig.~\ref{fig:3}~(a), where the exemplary modulated CW profile for $m=6.1\times10^{-3}$ is depicted together with the phase modulation profile (dashed gray line).

Next, the behavior of the TLS branch differs significantly with increasing $m$. Upon applying periodic phase modulation, the original unmodulated TLS branch (dark gray in the panel (i)) breaks up, leaving behind isolated loops of infinitesimal size for each point on the original TLS branch. These isolas are bounded by two fold bifurcations. However, as $m$ increases, the isolas immediately start to grow in width and reconnect with each other. This process of growth and reconnection is illustrated in Fig.~\ref{fig:3}~(ii)-(vi), with markers indicating points on the original CW (open circles) and TLS branches for $m=0$ in panel (ii) that are tracked as $m$ increases across the panels.

In particular, in Fig.~\ref{fig:3}~(iii), for a small modulation amplitude of $m=0.4\times10^{-3}$, several infinitesimal isolas have already been reconnected. At this stage, four isolas are visible (in yellow and light brown), marked at the corresponding fold bifurcations by crosses, squares, and triangles. One loop is marked with a diamond and a star; this loop is the result of two previously reconnected loops. In panel (iv), at $m=1\times10^{-3}$, the isolas marked with squares and triangles have merged to form an extended isola (orange), whereas the isola marked with crosses grows (dark brown).
Further increasing $m$ to $1.5\times10^{-3}$ brings the isolated loop (dark blue) close to the deformed CW branch (see panel (v), triangle marker), indicating the modulation amplitude at which the isolated loops reconnect with the deformed CW branch. At higher values of $m$, such as $m=3.2\times10^{-3}$ (green) and $6.1\times10^{-3}$ (purple), as shown in Fig.~\ref{fig:3}(vi), the isolas have fully reconnected with the deformed CW branch, restoring the typical collapsed snaking structure with alternating stability regions.
\begin{figure}
\includegraphics{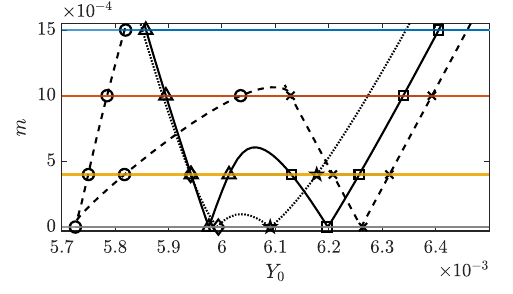} 
\caption{\label{fig:4} 
Two-parameter continuation of different fold points marked by black markers in Fig.~\ref{fig:3}~(ii)-(v) in the $(Y_0,\,m)$ plane showing the growth and reconnection of the TLSs isolas and modulated CW branches. Parameters are the same as in Fig.~\ref{fig:3}.
}
\end{figure}

The results presented in Fig.~\ref{fig:3}~(i)-(v) for small $m$ values can be summarized in a two-parameter bifurcation diagram in the $(Y_0,\,m)$-plane, see Fig.~\ref{fig:4}. Here, we follow the fold bifurcations bounding the isolas and the upper CW fold (cf. black markers in Fig.~\ref{fig:3}) and can analyze how they are connected as $m$ increases, illustrating the expanding and reconnecting behavior of the TLSs isolas and deformed CW branches. Here, horizontal cuts in gray, yellow, orange and blue correspond to $m$ values presented in Fig.~\ref{fig:3}~(i)-(v), respectively.

For instance, the TLSs isolas marked by triangles and squares each originate from two points on the stable part of the TLS branch at $m=0$ (cf. Fig.~\ref{fig:3}~(ii)). As $m$ increases, the created isolas expand in width and reconnect at the fold points facing each other. The continuation of these fold bifurcations forms a W-shaped curve in the $(Y_0,\, m)$ plane (cf. solid black line in Fig.~\ref{fig:4}), where the reconnection occurs as $m$ increases beyond the middle arc of the W-shape. After this point, the isolas merge into a single large loop. This reconnection pattern appears also for other fold pairs ranging from the diamond and star markers (cf. dotted line), to the folds marked by TLSs crosses and CW circles (dashed line). In this way, each point on the original TLS branch for $m=0$ becomes a tiny isola at small $m$, which then begins reconnecting with other isolas to form larger ones as $m$ increases, cf. Fig.~\ref{fig:3}~(ii)-(v). For large $m$, this results in a unified branch that resembles a slanted version of the original TLS branch (cf. Fig.~\ref{fig:3}~(i),\,(vi)).

The profiles of the modulated dark TLS solutions corresponding to these isolated loops are displayed in Fig.~\ref{fig:3}~(b)-(d). Here, the profile colors correspond to colors of the corresponding branches in Fig.~\ref{fig:3}~(i)-(vi). For small $m$ values, (e.g., the yellow profile in panel (b)), they closely resemble the original case at $m=0$ (dark gray), with the primary difference being a slight modulation of the background intensity due to the periodic phase modulation, similar to the modulated CW solutions, cf. Fig.~\ref{fig:3}~(a). As $m$ increases, the curvature of the high-intensity CW background on which the TLS solutions live becomes more pronounced, as shown in Fig.~\ref{fig:3}~(c) and (d), consistent with the analysis of the CW branch.

\subsection{Hermite--Gauss modes}
\label{sec:HG_modes}
\begin{figure}
\includegraphics{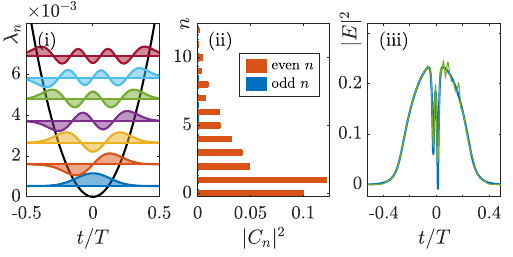}
\caption{\label{fig:5}
(i) First seven HG modes within a parabolic potential. (ii) Mode energy distribution associated with the TLS shown in blue in panel (iii). (iii) Comparison of TLS (blue) and superposition of HG modes weighted with the energy distribution from panel (ii) (green). The system parameters are as in Fig.~\ref{fig:3} and $m=0.1$.}
\end{figure}

As discussed above, in the UFL, where $\eta\approx1$ and $\tilde{\varphi}$ is small, the dynamics of the KGTI system~\eqref{eq:whole_KGTI} can be well described by the LLE with TOD~\eqref{eq:LLE}. Considering the situation of one pulse per modulation period, we approximate the weak modulation $\tilde{\varphi}=\tilde{\varphi}(\sigma)$ as a parabolic potential, yielding
\begin{equation}
    \tilde{\varphi}(\sigma)=\frac{m\,\omega{{}^2}}{2}\sigma{{}^2}+\Delta\varphi\,,
\end{equation}
where $\Delta\varphi=m+\varphi_{0}-2\arctan(\delta)$. Hence, for $\delta=1/\sqrt{3}$, where the TOD coefficient $\beta_3=0$ and for the cold cavity, where the effects of the nonlinearity and dissipation are neglected, the LLE Eq.~\eqref{eq:LLE} reduces to a linear quantum harmonic oscillator
\begin{equation}
-i\,\partial_{\theta}E=\left(C\sigma^{2} + \Delta\varphi\right)E + B\,\partial_{\sigma}^2E\,,
\label{eq:lin_LLE_2}
\end{equation}
where $B=-\beta_2/2$ and $C=m\omega^2/2$. The ansatz $E(\sigma,\theta)=E_s(\sigma)\,e^{-i\lambda_n\theta}$, where $\lambda_n$ being the frequency of the eigenmode, leads to a singular Sturm-Liouville problem similar to that studied in~\cite{GMJ-PRR-24}
$$
\left(\lambda_n+\Delta\varphi+C\sigma^2+B\partial_{\sigma}^2\right)E_s=0,\quad \lim_{x\rightarrow\pm\infty}E_s=0.
$$
For $BC<0$ the solutions of its homogeneous part are the well-known Hermite--Gauss (HG) modes~\cite{GMJ-PRR-24,SWPR-OL-23,SPRF-OL-22} $\psi_n(\sigma/d)$ with the width
\begin{equation}
    d^2 = \sqrt{-\frac{B}{C}} = \sqrt{\frac{\beta_2}{m\omega^2}}\,.\label{eq:hg_width}
\end{equation}
The corresponding eigenfrequencies $\lambda_n$ read
\begin{equation}
    \lambda_{n}=\frac{B}{d^2}(2n+1)-\Delta\varphi\,,
\end{equation}
where $n\in\mathbb{N}$ is the modal index.

In Fig.~\ref{fig:5}~(i) we show the first seven HG modes inside a parabolic potential that are located at the corresponding $\lambda_n$.
Since the HG modes form a complete orthogonal basis, solutions $E(\sigma,\theta)$  of the linear Eq.~\eqref{eq:lin_LLE_2} can be expanded in HG modes as
\begin{equation}
    E(\sigma,\theta) = \sum_n C_n(\theta) e^{i\lambda_n \theta} \psi_n(\sigma)\,,
\end{equation}
where
\begin{equation}
    C_{n}(\theta)e^{i\lambda_{n}(\theta)} = \int_{-\infty}^{\infty}E(\sigma,\theta)\psi_{n}(\sigma)\mathrm{d}\sigma 
\end{equation}
is the complex amplitude. The absolute value $|C_n(\theta)|^2$ gives the mode energy distribution~\cite{SPRF-OL-22}. Although the HG approximation was made assuming a lossless linear cavity, we calculated the mode energy distribution for a TLS of the full DAE model~\eqref{eq:whole_KGTI}, obtained for the same parameters as TLSs in Fig.~\ref{fig:3}, but for $m=0.1$ (cf. also Fig.~\ref{fig:6}~(ii)). The results are presented in Fig.~\ref{fig:5}~(ii), (iii).

In Fig.~\ref{fig:5}~(ii) one can see the mode energy distribution corresponding to the TLS in Fig.~\ref{fig:5}~(iii). Here, only even modes contribute significantly to the pulse shape. The mode energies of the odd modes are negligible and not visible in the bar diagram. This underlines the fact that the detuning $\delta$ was chosen such that $\beta_3=0$ resulting in almost symmetrical temporal profile and higher-order odd terms, such as the fifth-order terms, contribute minimally. Figure~\ref{fig:5}~(iii) compares the original TLS profile (blue) with the superposition of HG modes weighted by the calculated mode energies $|C_n|^2$ (green). The good agreement demonstrates that the cavity's dynamics can be well described by HG modes for the chosen parameters in the UFL.

%

\subsection{Resonance curves}
\begin{figure}
\includegraphics{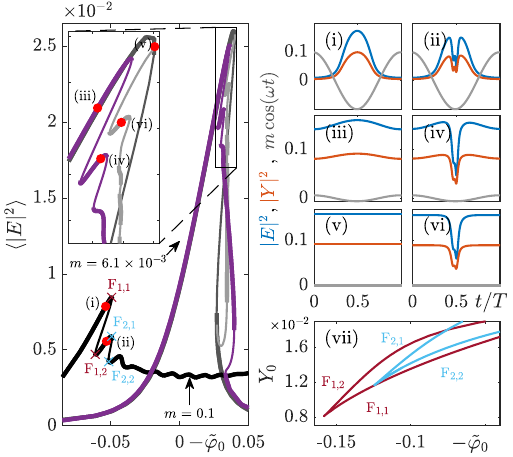}
\caption{\label{fig:6}
Left: Bifurcation diagram showing the integrated intensity $\langle|E|^2\rangle$
vs.~the effective unmodulated feedback phase ${\tilde{\varphi}_0=\varphi_0-2\arctan\delta}$ for $Y_0=0.0066$ for three different values of the modulation amplitude $m$. The CW and TLS for $m=0$ are shown in dark and light gray, respectively, while the resonance curves for $m=\{6.1\times10^{-3},10^{-1}\}$ are depickted in purple and black, respectively.
Panels (i) - (vi) show the exemplary profiles corresponding to the red markers in the left panel, where the microcavity field intensity $|E|^2$, the external cavity field intensity $|Y|^2$ and the phase modulation are presented in orange, blue and gray, respectively. (vii) shows two-parameter continuations in $-\tilde{\varphi}_0$ and $Y_0$ of two pairs of fold bifurcations marked in the left panel with the light blue and dark red crosses. Other parameters are as in Fig.~\ref{fig:3}.
}
\end{figure}

In the context of the LLE~\eqref{eq:LLE}, the unmodulated effective feedback phase $\tilde{\varphi}_0=\varphi_0-2\arctan\delta$ of the KGTI system~\eqref{eq:whole_KGTI} plays the role of the cavity detuning. In Ref.~\onlinecite{SWPR-OL-23}, the impact of the parabolic potential on the dynamics of the LLE without TOD was studied, where the cavity detuning was used as a bifurcation parameter. To compare our results of the full DAE system~\eqref{eq:whole_KGTI} in the UFL and for $\beta_3=0$ with the results obtained in Ref.~\onlinecite{SWPR-OL-23}, we perform a bifurcation analysis of the system~\eqref{eq:whole_KGTI} for fixed $Y_0$ and the same parameters as in Fig.~\ref{fig:3}, using $-\tilde{\varphi}_0$ as a control parameter to have the same sign convention as in the LLE in Ref.~\onlinecite{SWPR-OL-23}

Our results are presented in Fig.~\ref{fig:6}, where the left panel shows the resonances of the nonlinear cavity for three different $m=(0,\, 6.1\times10^{-3},\,0.1)$. In particular, the dark gray curve shows a tilted resonance peak for the CW solution at $m=0$~\cite{SJG-OL-24}. Similar to the branches in injection $Y_0$ (cf. Fig.~\ref{fig:3}), the TLS solutions (light gray) bifurcate from the CW and build a collapsed snaking branch (see also the zoom in the left panel of Fig.~\ref{fig:6}). Two exemplary profiles at the points marked by red circles are presented in Fig.~\ref{fig:6}~(v) and (vi), where $|E|^2$, $|Y|^2$ and modulation are shown in blue, orange, and gray, respectively.

As the modulation amplitude is increased to $m=6.1\times10^{-3}$, the modulated CW and TLS branches have been already reconnected as discussed in Sec.~\ref{sec:bif_ufl} (cf. Fig.~\ref{fig:3}), and the resulting single resonance curve is presented in Fig.~\ref{fig:6} in purple. One can see, that the branch becomes slightly tilted but largely retains its original structure, while modulated TLSs do not bifurcate anymore from the modulated CW, but are glued to a single branch.

In the intensity profiles, this increase in modulation amplitude induces a modulation in the high-intensity background, as shown in Fig.~\ref{fig:6}~(iii) and (iv) (cf.~similar profiles in Fig.~\ref{fig:3}~(a),~(d)). In the light of the discussion of Sec.~\ref{sec:HG_modes}, the modulated profile in Fig.~\ref{fig:6}~(iii) can be interpreted as a weakly nonlinear and dissipative HG mode with the width $d$, that is inversely proportional to $m$ (cf. Eq.~\eqref{eq:hg_width}), is too large to fit into the cavity size. A first dark modulated TLS in Fig.~\ref{fig:6}~(iii) seats on the top of the mode, but resembles its original shape for $m=0$ (cf. Fig.~\ref{fig:6}~(vi)).

Further increasing the modulation amplitude to a large value of $m=0.1$ results in a fundamentally different behavior of the modulated resonance branch, see the black curve in the left panel of~Fig.~\ref{fig:6}. The collapsed snaking structure is completely overturned, with the windings of the snaking branch transforming into individual resonance peaks. These peaks correspond to the weakly nonlinear and dissipative HG modes, which dominate the dynamics at high modulation amplitudes $m$, see two exemplary profiles in panels (i) and (ii). In particular, Fig.~\ref{fig:6}~(i) shows a profile resembling a HG fundamental mode with the width now fitting in to the cavity for large $m$ values, cf. Eq.~\eqref{eq:hg_width}. Fig.~\ref{fig:6}~(ii) can be interpreted as a tiny dark TLS, sitting on the top of the mode, but can be also seen as a weakly nonlinear and dissipative combination of HG modes, where the mode $\psi_2(\sigma)$ is dominant, cf. Fig.~\ref{fig:5}~(iii). Note that the shape of the observed resonance curves found in Fig.~\ref{fig:6} agrees very well with the results presented in Ref.~\onlinecite{SWPR-OL-23,SPRF-CSF-23} for the LLE equation.

Increasing the injection $Y_0$, more energy gets introduced into the cavity, raising the intensities and the Kerr effect, further tilts the resonance peak. This results in the expansion of the bistability range~\cite{SWPR-OL-23}. A two-parameter continuation in $Y_0$ and $-\tilde{\varphi}_0$ of the two pairs of fold bifurcations, marked in red and light blue in Fig.~\ref{fig:6} is presented in Fig.~\ref{fig:6}~(vii). Depending on the value of $Y_0$ (with $\tilde{\varphi}_0$ fixed), the number of fold bifurcations can be controlled. Note that the dynamics of folds agrees well with the fold evolution found in the LLE~\cite{SWPR-OL-23}.
%

\section{\label{sec:full} Beyond the uniform field limit}

While the analysis in the  UFL is a feasible starting point for the analysis, the description of the modulated KGTI system~\eqref{eq:whole_KGTI} is not limited to this regime, while the mean-field models like LLE are not always able to account for the complete range of complex dynamics outside the UFL.
In particular, in what follows, we choose a parameter set far away from UFL~\cite{SJG-OL-24} as used in Figs.~\ref{fig:1},~\ref{fig:2} and perform a bifurcation analysis.


\subsection{Bifurcation analysis beyond the UFL limit}
\begin{figure}
\includegraphics{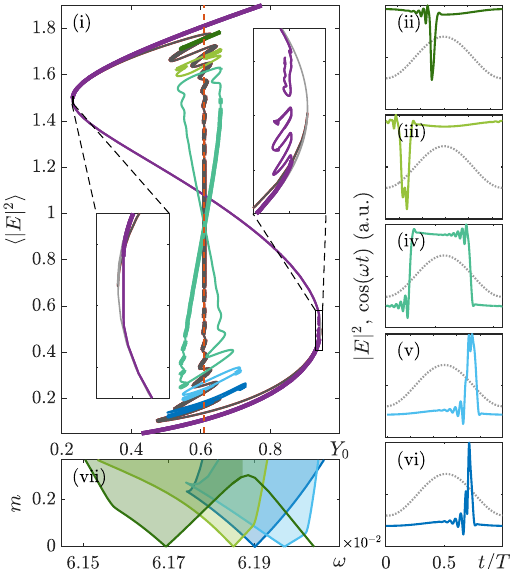}
\caption{\label{fig:7} 
(i) Bifurcation diagram showing the integrated intensity $\langle|E|^2\rangle$ as a function of the injection $Y_0$ for three values of $m$ at fixed $\omega=0.0618$. CW and TLS branches for $m=0$ are shown in light and dark gray, respectively. The CW branch for $m=0.01$ is depicted in purple. The insets magnify the fold bifurcation points of the CW branch. Dark blue, light blue, turquoise, light and dark green branches corresponding to the TLSs for $m=0.22$. Panels (ii) - (vi) show the profiles corresponding to the different TLSs isolas at $Y_0=0.609$ (dashed orange line) together with the phase modulation (gray dotted). (vii) Arnold tongues corresponding to the TLSs in panels (ii) - (vi) for $Y_0=0.609$. Other parameters as in Fig.~\ref{fig:1}.
}
\end{figure}

In general, in the normal dispersion regime, the unmodulated system~\eqref{eq:whole_KGTI} exhibits a bistable CW response. There, both bright and dark TLSs appear via the locking of domain walls connecting the high and low CWs. Due to the oscillatory tails induced by the cavity dispersion, these TLSs can interlock at multiple equilibrium distances leading to a rich ensemble of multistable bright and dark solutions~\cite{SPV-OL-19,SJG-OL-22,SJG-OL-24}. This situation for the fixed modulation frequency $\omega$ is presented in Fig.~\ref{fig:7}~(i) in light and dark gray for CW and TLS branches, respectively.

In the presence of a modulation with $m = 0.01$, the CW branch (purple), for the most part, is identical to the CW branch at $m = 0$, see Fig.~\ref{fig:7}~(i). However, as discussed above in Sec.~\ref{sec:bif_ufl}, the phase modulation causes the CW to deform at the fold bifurcation points (see the insets in Fig.~\ref{fig:7}~(i) for a closer comparison). The deformation at the upper fold bifurcation point is minor, but at the lower fold bifurcation point, the deformations are significantly more pronounced. Here, the CW branch has already split and reconnected with some complex and irregular unstable structures. Nevertheless, all these deformations of the CW branch are located close to the folds and do not result in any observable behavior in numerical time simulations. Hence, we do not show the CW dynamics for larger values of $m$ and we will restrict our study of the CW branch to this small modulation amplitude.

For the TLSs, at a finite modulation strength we observe that the branch fragments into a stack of isolated structures of different sizes, similar to the dynamics observed in the UFL limit. The resulting bifurcation diagram for $m=0.22$ is shown in Fig.~\ref{fig:7}~(i) in dark blue, light blue, turquoise, light and dark green. Panels (ii) - (vi) illustrate the corresponding TLS profiles with the same color together with the phase modulation (gray) corresponding to the injection value marked with the orange vertical dashed line in Fig.~\ref{fig:7}~(i). One can see the the resulting isolas are still well-centered around the same Maxwell line as for $m=0$, although the vertical branch part for $m=0$ becomes an extended loop.

We note that the corresponding profiles of the TLSs in panels (ii)-(vi) are very similar to those obtained in the KGTI system without phase modulation~\cite{SJG-OL-22,SJG-OL-24}, but all possessing different position in the cavity with respect to the presented modulation. The phase modulation only slightly modifies the low- or high-intensity CW background values. This observation is significantly different from what was obtained in the UFL limit in Sec.~\ref{sec:ufl}, where the dynamics was enslaved by the HG modes (cf. Figs.~\ref{fig:3},\,\ref{fig:6}). However, outside the UFL limit and in the presence of TOD, the weakly nonlinear and dissipative HG modes are not solutions of the system in question and, hence, do not influence the occurring dynamics.

\subsection{Synchronization of TLSs}

Now, the question arises, whether one can control how many (and which) TLS isolas  appear for different modulation parameters $m$ and $\omega$. To answer this question we change the perspective again and consider the appearance of the TLS in the presence of the modulation as synchronization process as was discussed in Sec.~\ref{sec:synchronization} (cf. Fig.~\ref{fig:2}). Hence, we fix the injection $Y_0$ to the value corresponding to panels Fig.~\ref{fig:6}~(ii)-(vi), and perform the continuation of the TLSs solutions of different widths in the modulation frequency $\omega$. The folds of the resulting synchronized regions are then continued in the modulation amplitude $m$. The result is presented in Fig.~\ref{fig:6}~(vii). Here, the color of resulting Arnold tongues correspond to the colors of TLSs in Fig.~\ref{fig:6}~(ii)-(vi).

We find that different TLSs of different widths can be synchronized. Here, the resulting Arnold tongues possess different widths and are located around different frequencies. Depending on $m$, several tongues can overlap, leading to the multistability of the synchronized TLSs. In particular, at the point $(\omega, m) = (0.0618, 0.22)$, all the tongues overlap, leading to to bifurcation structure corresponding to Fig.~\ref{fig:6}~(i).
Note that the dark green tongue, hosting the synchronized dark TLS, exhibits a different behavior from the others, with the right-side border curving downward and reducing back to $m=0$. However, on the right side of the bend, all solutions are unstable. Therefore, this curious part of the Arnold tongue does not influence the resulting dynamics.

%

\subsection{Position of modulated TLSs}
\begin{figure}
\includegraphics{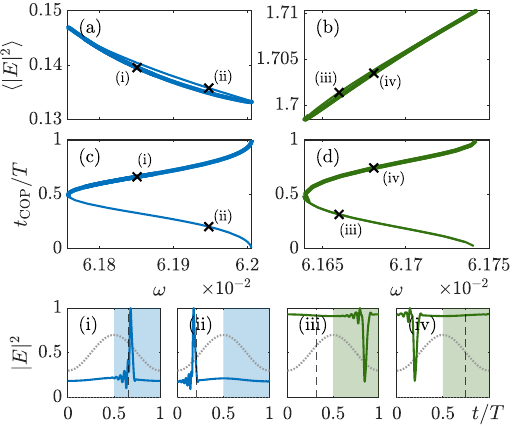}
\caption{\label{fig:8} (a), (b) Branch of bright (dark) TLS showing the integrated-intensity of the $E$-field as a function of the modulation frequency $\omega$.
(c), (d): Same branches using the scaled center-of-pulse $t_\mathrm{COP}$ norm. Panels (i) - (iv) show the pulse profiles at the positions, marked with black crosses.
Black dashed lines indicate the respective value of the center-of-pulse norm. The phase modulation is indicated with the dotted gray line. The shaded areas corresponds to the stable cavity sections. Parameters are as in Fig.~\ref{fig:7}.
}
\end{figure}

The results presented in Fig.~\ref{fig:7} reveal that the modulated TLSs possess different positions in the modulation landscape (cf. also Figs.~\ref{fig:1},~\ref{fig:2}) and are not located at the potential minima as in the UFL case (cf. Fig.~\ref{fig:6}). Numerical time simulations of the DAE~\eqref{eq:whole_KGTI} have verified that in the synchronized case TLSs consistently travel to a specific position within the phase modulation, depending on their shape and the values for $\omega$ and $m$.
%
%
To study the pulse positions, we implement a so-called \textit{center-of-pulse} (COP) norm $t_\mathrm{COP}$, corresponding to the center of mass of a pulse on a circular domain \cite{BB-JGT-08}, to account for the periodicity of the solutions. First, we focus on the first bright TLS, previously depicted in dark blue in Fig.~\ref{fig:7}~(i), (vi). A continuation in $\omega$ for the fixed $Y_0$ plotted with the integrated intensity measure forms an isolated loop, as shown in Fig.~\ref{fig:8}~(a), with the upper half unstable and the lower part stable, similar to Fig.~\ref{fig:2}~(vi). In the corresponding COP norm in Fig.~\ref{fig:8}~(c), we observe a periodic, cosine-like variation of the position $t_\mathrm{COP}$ throughout the isola. The stability changes precisely at the fold points of the isola and not in between. The positions of the pulses at the folds coincides with the extrema of the phase modulation. For the stable section, the bright pulse lives on the rising edge of the phase modulation, while for the unstable section, the pulse is situated on the falling edge. This is illustrated in Fig.~\ref{fig:8}~(i) and (ii) with the colored background marking the stable cavity parts. Here, the black dashed line shows the respective value of $t_\mathrm{COP}$.
For the dark TLS (dark green in \ref{fig:7}~(i), (ii)), Fig.~\ref{fig:8}~(b) and (d) are calculated analogously and show the same trend. However, the position of the dashed line for the profiles, presented in Fig.~\ref{fig:8}~(iii) and (iv) may seem counterintuitive. Nonetheless, since the calculation of $t_\mathrm{COP}$ is analogous to determining the center of mass on a periodic domain, this position precisely indicates where the highest "mass" — that is, the most intensity — is located within one period. Therefore, it is important to distinguish between the visual position of the dark TLS and the position given by $t_\mathrm{COP}$. Note that both bright and dark TLSs, are stable when $t_\mathrm{COP} \in [0.5, 1]$. However, if one considers the visual position of the dark TLS, the dip that forms the dark TLS must be within the interval $[0, 0.5]$ for the TLS to be stable.

The reason behind the observed behavior for both bright and dark TLSs lies in the interaction between the TLS and the phase modulation. The rising edge of the phase modulation supports the transition to the high-intensity CW state, stabilizing the bright TLS. Conversely, the falling edge hinders this transition, leading to the pulse instability. For the dark TLS, the stable region is where the falling edge of the phase modulation facilitates the transition to the low-intensity state, while the rising edge causes instability. Similarly, wider TLS, such as those in Fig.~\ref{fig:7}~(iv), are stable when the connection from the low to high-intensity CW solution is on the rising edge of the phase modulation and the connection from high to low intensity is on the falling edge. Reversing these positions, or shifting the pulse by half a period, would render the pulse unstable. This behavior is analogous to that of the narrower TLS.

\subsection{Effective equation of motion}
\begin{figure}
\includegraphics{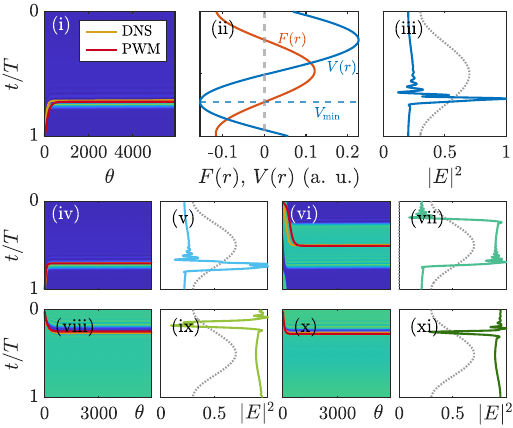}
\caption{\label{fig:9} (i) Time trace of the first bright TLS in two-time representation. The yellow line marks the position of the pulse determined from the time trace and the red line show the pulse position as determined with the EOM~\eqref{eq:KGTI_one_eq}. (ii) Force $F(r)$ (orange) and the potential $V(r)$ (blue) as determined from the Eq.~\eqref{eq:KGTI_one_eq}. The potential minimum $V_\mathrm{min}$ is marked with the blue dashed line. (iii) TLS profile corresponding to the time trace shown in (i) together with the phase modulation indicated by the dotted gray line.
Panels (iv)-(xi) show the results of the EOM for TLS of different width. Parameters are as in Fig.~\ref{fig:7}.
} 
\end{figure}

As discussed above, we observed that TLSs in the KGTI system~\eqref{eq:whole_KGTI} with phase modulation, settle into specific positions within the phase modulation, depending on the modulation parameters $(\omega,\, m)$. Specifically, near the edge of the synchronization region, pulses tend to localize at the extrema of the phase modulation. However, for intermediate parameter values, precisely predicting the pulse positions becomes challenging besides the general periodic position shape presented in Fig.~\ref{fig:8}~(c),(d).

Hence, we derive a so-called potential-well model, i.e., an effective equation of motion (EOM) for the pulse position within the potential, viewing the TLS pulse as a particle moving within a potential landscape. In particular, we consider the system in question without modulation and assume the influence of the modulation on the system is only a small perturbation. Similar modeling approaches have successfully explained phenomena such as e.g., front propagation in one- and two-dimensional spatially modulated media~\cite{HER-PRE-10}, delay-induced depinning of localized structures in the inhomogeneous Swift-Hohenberg equation~\cite{TST-PRE-17}, Lugiato-Lefever equation with inhomogeneous injection ~\cite{TFHP-PRA-19}, the interaction of topological solitons in the delayed Adler equation \cite{MJG-Chaos-20}, arrested motion of isolated domain walls~\cite{JAH-PRL-15}, localized passively-mode-locked pulses in a time-dependent current landscape~\cite{CJM-PRA-16} and desynchronization of localized states in Kerr cavity with pulsed injection~\cite{DHT-OL-24}, to name just a few.

We start by rewriting the full DAEs~\eqref{subeq:KGTI_E} and \eqref{subeq:KGTI_Y} in vectorial form:
\begin{equation}
    \mathbf{M}\dot{\mathbf{\psi}}(t)=\mathbf{N}_m{\left[\mathbf{\psi}(t),\mathbf{\psi}(t-\tau),\,t\right]}
    \label{eq:KGTI_one_eq}
\end{equation}
with $\mathbf{\psi=\left(\mathrm{Re}(E),\,\mathrm{Im}(E),\,\mathrm{Re}(Y),\,\mathrm{Im}(Y) \right)^T}$, $\mathbf{M}$ a mass matrix and $\mathbf{N}_m$ is a r.h.s. nonlinear operator.
We denote with $\mathbf{\psi}_{0}=\mathbf{\psi}_{0}(t)$ the unperturbed solutions of the system~\eqref{eq:KGTI_one_eq} without phase modulation, i.e. for $m=0$. Further, we assume that the periodic phase modulation induces a small perturbation, primarily causing a temporal shift of the pulse without altering its overall shape.

Thus, we employ the ansatz $\mathbf{\psi}\left(t\right)=\mathbf{\psi}_{0}\left(t-r(t)\right)$ which represents the original TLS solutions shifted to some slow varying position $r(t)$, while neglecting any shape deformations. With this ansatz and projecting onto the Goldstone mode of the adjoint system without phase modulation $\mathbf{v}_0$, we arrive to the effective equation of motion for the pulse position $r(t)$:
\begin{equation}
     \dot{r}(t) = 1 - \frac{\big\langle \mathbf{v}_0{\left(t'\right)} \big| \mathbf{N}_m{\big[\mathbf{\psi}_{0}{\left(t'\right)},\mathbf{\psi}_{0}\big(t'-\tau \big),t'+r(t)\big]} \big\rangle}{\left\langle \mathbf{v}_0{\left(t'\right)} \big| \mathbf{N}_{0}{\left[\mathbf{\psi}_{0}(t'),\mathbf{\psi}_0(t'-\tau)\right]} \right\rangle} .
     \label{eq:diff_eq_r}
\end{equation}
Here, the notation $\langle\cdot|\cdot\rangle$ stands for the scalar product defined by integration over the full domain and $\mathbf{N}_{0}=\mathbf{N}_{m=0}$. Note that the argument of the delayed fields simplifies under the assumption that changes in $r$ are small over one round trip, i.e. $r(t)\approx r(t-\tau)$. The denominator acts as a normalization and only depends on the unmodulated system.  The right-hand side of the effective equation of motion can be interpreted as a force $F(r)$ depending on the position $r$.

Figure~\ref{fig:9}~(ii) presents an example of the force $F(r)$ (orange) and the corresponding potential $V(r)$, defined as $V(r) = -\int_{0}^r F(r')\mathrm{d}r'$ (blue), evaluated for the bright TLS that is shown in panel (iii) together with the phase modulation (cf. Fig.~\ref{fig:7}~(vi) and Fig.~\ref{fig:8}~(i)). One can see, that the force is a periodic function, whereas its roots correspond to the extrema of the potential $V(r)$.

Figure~\ref{fig:9}~(i) shows a time trace of the first bright TLS in two-time representation. Here, the yellow line marks the position of the pulse determined from the time trace and the red line indicates the pulse position as determined with the potential-well model Eq.~\eqref{eq:diff_eq_r}. One can see that the pulse from a direct time integration, shown in a two-time diagram, settles nicely in the minimum of the potential $V_\mathrm{min}$, cf. panel (ii), similar to the behavior of a ball rolling down a hill in the potential created by gravity.
%
%
Moreover, even in the transient phase the pulse position determined in direct numerical simulation and in the EOM are in a good agreement.

Additionally, Fig.~\ref{fig:9}~(iv)-(xi) show the comparison of the EOM and the pulse position evaluated from the full time integration for TLSs of different width (cf. Fig.~\ref{fig:7}). In all the cases, the EOM nicely predicts the final position TLS position, having some deviations in the transient area, where the pulse shape is changing.

\section{Conclusions}
\label{sec:summary}
In conclusion, we have theoretically analyzed the influence of phase modulation on the dynamics of TLSs in the KGTI system operating in the normal dispersion regime. We have shown that phase modulation induces a synchronization mechanism between the external forcing and the intrinsic pulse dynamics, giving rise to a finite locking region in which TLSs are locked to the imposed modulation. A detailed bifurcation analysis of the full delay algebraic model in the uniform field limit revealed that the dynamics of dark localized structures are governed by weakly nonlinear dissipative Hermite--Gauss modes, leading to a rich bifurcation scenario. Beyond the uniform field limit, both bright and dark modulated TLSs were found to stable in a certain parameter range and to occupy distinct equilibrium positions determined by the modulation profile. An effective equation of motion for the TLSs positions was derived and shown to be in good agreement with the full delay-algebraic model, providing a reduced and physically transparent description of the observed dynamics.

\begin{acknowledgments}
J.J. acknowledges the financial support of the project KEFIR/AEI/10.13039/501100011033/ FEDER, UE. J.J. and S.V.G. acknowledge the financial support of the project KOGIT, Agence Nationale de la Recherche (ANR-22-CE92-0009), Deutsche Forschungsgemeinschaft (DFG), Germany via Grant Nr. 505936983 and Nr. 524947050.
\end{acknowledgments}

\section*{Data Availability Statement}

The data that support the findings of this study are available from the corresponding author upon reasonable request.

\section*{References}

\begin{thebibliography}{59}%
\makeatletter
\providecommand \@ifxundefined [1]{%
 \@ifx{#1\undefined}
}%
\providecommand \@ifnum [1]{%
 \ifnum #1\expandafter \@firstoftwo
 \else \expandafter \@secondoftwo
 \fi
}%
\providecommand \@ifx [1]{%
 \ifx #1\expandafter \@firstoftwo
 \else \expandafter \@secondoftwo
 \fi
}%
\providecommand \natexlab [1]{#1}%
\providecommand \enquote  [1]{``#1''}%
\providecommand \bibnamefont  [1]{#1}%
\providecommand \bibfnamefont [1]{#1}%
\providecommand \citenamefont [1]{#1}%
\providecommand \href@noop [0]{\@secondoftwo}%
\providecommand \href [0]{\begingroup \@sanitize@url \@href}%
\providecommand \@href[1]{\@@startlink{#1}\@@href}%
\providecommand \@@href[1]{\endgroup#1\@@endlink}%
\providecommand \@sanitize@url [0]{\catcode `\\12\catcode `\$12\catcode
  `\&12\catcode `\#12\catcode `\^12\catcode `\_12\catcode `\%12\relax}%
\providecommand \@@startlink[1]{}%
\providecommand \@@endlink[0]{}%
\providecommand \url  [0]{\begingroup\@sanitize@url \@url }%
\providecommand \@url [1]{\endgroup\@href {#1}{\urlprefix }}%
\providecommand \urlprefix  [0]{URL }%
\providecommand \Eprint [0]{\href }%
\providecommand \doibase [0]{http://dx.doi.org/}%
\providecommand \selectlanguage [0]{\@gobble}%
\providecommand \bibinfo  [0]{\@secondoftwo}%
\providecommand \bibfield  [0]{\@secondoftwo}%
\providecommand \translation [1]{[#1]}%
\providecommand \BibitemOpen [0]{}%
\providecommand \bibitemStop [0]{}%
\providecommand \bibitemNoStop [0]{.\EOS\space}%
\providecommand \EOS [0]{\spacefactor3000\relax}%
\providecommand \BibitemShut  [1]{\csname bibitem#1\endcsname}%
\let\auto@bib@innerbib\@empty
\bibitem [{\citenamefont {Diddams}(2010)}]{D-JOSAB-10}%
  \BibitemOpen
  \bibfield  {author} {\bibinfo {author} {\bibfnamefont {S.~A.}\ \bibnamefont
  {Diddams}},\ }\href {\doibase 10.1364/JOSAB.27.000B51} {\bibfield  {journal}
  {\bibinfo  {journal} {J. Opt. Soc. Am. B}\ }\textbf {\bibinfo {volume}
  {27}},\ \bibinfo {pages} {B51} (\bibinfo {year} {2010})}\BibitemShut
  {NoStop}%
\bibitem [{\citenamefont {Pasquazi}\ \emph {et~al.}(2018)\citenamefont
  {Pasquazi}, \citenamefont {Peccianti}, \citenamefont {Razzari}, \citenamefont
  {Moss}, \citenamefont {Coen}, \citenamefont {Erkintalo}, \citenamefont
  {Chembo}, \citenamefont {Hansson}, \citenamefont {Wabnitz}, \citenamefont
  {Del'Haye}, \citenamefont {Xue}, \citenamefont {Weiner},\ and\ \citenamefont
  {Morandotti}}]{PPR-PR-18}%
  \BibitemOpen
  \bibfield  {author} {\bibinfo {author} {\bibfnamefont {A.}~\bibnamefont
  {Pasquazi}}, \bibinfo {author} {\bibfnamefont {M.}~\bibnamefont {Peccianti}},
  \bibinfo {author} {\bibfnamefont {L.}~\bibnamefont {Razzari}}, \bibinfo
  {author} {\bibfnamefont {D.~J.}\ \bibnamefont {Moss}}, \bibinfo {author}
  {\bibfnamefont {S.}~\bibnamefont {Coen}}, \bibinfo {author} {\bibfnamefont
  {M.}~\bibnamefont {Erkintalo}}, \bibinfo {author} {\bibfnamefont {Y.~K.}\
  \bibnamefont {Chembo}}, \bibinfo {author} {\bibfnamefont {T.}~\bibnamefont
  {Hansson}}, \bibinfo {author} {\bibfnamefont {S.}~\bibnamefont {Wabnitz}},
  \bibinfo {author} {\bibfnamefont {P.}~\bibnamefont {Del'Haye}}, \bibinfo
  {author} {\bibfnamefont {X.}~\bibnamefont {Xue}}, \bibinfo {author}
  {\bibfnamefont {A.~M.}\ \bibnamefont {Weiner}}, \ and\ \bibinfo {author}
  {\bibfnamefont {R.}~\bibnamefont {Morandotti}},\ }\href {\doibase
  https://doi.org/10.1016/j.physrep.2017.08.004} {\bibfield  {journal}
  {\bibinfo  {journal} {Physics Reports}\ }\textbf {\bibinfo {volume} {729}},\
  \bibinfo {pages} {1 } (\bibinfo {year} {2018})}\BibitemShut {NoStop}%
\bibitem [{\citenamefont {Tropper}\ \emph {et~al.}(2004)\citenamefont
  {Tropper}, \citenamefont {Foreman}, \citenamefont {Garnache}, \citenamefont
  {Wilcox},\ and\ \citenamefont {Hoogland}}]{tropper04}%
  \BibitemOpen
  \bibfield  {author} {\bibinfo {author} {\bibfnamefont {A.~C.}\ \bibnamefont
  {Tropper}}, \bibinfo {author} {\bibfnamefont {H.~D.}\ \bibnamefont
  {Foreman}}, \bibinfo {author} {\bibfnamefont {A.}~\bibnamefont {Garnache}},
  \bibinfo {author} {\bibfnamefont {K.~G.}\ \bibnamefont {Wilcox}}, \ and\
  \bibinfo {author} {\bibfnamefont {S.~H.}\ \bibnamefont {Hoogland}},\
  }\href@noop {} {\bibfield  {journal} {\bibinfo  {journal} {J. Phys. D: Appl.
  Phys.}\ }\textbf {\bibinfo {volume} {37}},\ \bibinfo {pages} {R75} (\bibinfo
  {year} {2004})}\BibitemShut {NoStop}%
\bibitem [{\citenamefont {Laurain}\ \emph {et~al.}(2010)\citenamefont
  {Laurain}, \citenamefont {Myara}, \citenamefont {Beaudoin}, \citenamefont
  {Sagnes},\ and\ \citenamefont {Garnache}}]{LMB-OE-10}%
  \BibitemOpen
  \bibfield  {author} {\bibinfo {author} {\bibfnamefont {A.}~\bibnamefont
  {Laurain}}, \bibinfo {author} {\bibfnamefont {M.}~\bibnamefont {Myara}},
  \bibinfo {author} {\bibfnamefont {G.}~\bibnamefont {Beaudoin}}, \bibinfo
  {author} {\bibfnamefont {I.}~\bibnamefont {Sagnes}}, \ and\ \bibinfo {author}
  {\bibfnamefont {A.}~\bibnamefont {Garnache}},\ }\href {\doibase
  10.1364/OE.18.014627} {\bibfield  {journal} {\bibinfo  {journal} {Opt.
  Express}\ }\textbf {\bibinfo {volume} {18}},\ \bibinfo {pages} {14627}
  (\bibinfo {year} {2010})}\BibitemShut {NoStop}%
\bibitem [{\citenamefont {Leo}\ \emph {et~al.}(2010)\citenamefont {Leo},
  \citenamefont {Coen}, \citenamefont {Kockaert}, \citenamefont {Gorza},
  \citenamefont {Emplit},\ and\ \citenamefont {Haelterman}}]{LCK-NAP-10}%
  \BibitemOpen
  \bibfield  {author} {\bibinfo {author} {\bibfnamefont {F.}~\bibnamefont
  {Leo}}, \bibinfo {author} {\bibfnamefont {S.}~\bibnamefont {Coen}}, \bibinfo
  {author} {\bibfnamefont {P.}~\bibnamefont {Kockaert}}, \bibinfo {author}
  {\bibfnamefont {S.}~\bibnamefont {Gorza}}, \bibinfo {author} {\bibfnamefont
  {P.}~\bibnamefont {Emplit}}, \ and\ \bibinfo {author} {\bibfnamefont
  {M.}~\bibnamefont {Haelterman}},\ }\href {\doibase 10.1038/nphoton.2010.120}
  {\bibfield  {journal} {\bibinfo  {journal} {Nat Photon}\ }\textbf {\bibinfo
  {volume} {4}},\ \bibinfo {pages} {471} (\bibinfo {year} {2010})}\BibitemShut
  {NoStop}%
\bibitem [{\citenamefont {Herr}\ \emph {et~al.}(2014)\citenamefont {Herr},
  \citenamefont {Brasch}, \citenamefont {Jost}, \citenamefont {Wang},
  \citenamefont {Kondratiev}, \citenamefont {Gorodetsky},\ and\ \citenamefont
  {Kippenberg}}]{HBJ-NAP-14}%
  \BibitemOpen
  \bibfield  {author} {\bibinfo {author} {\bibfnamefont {T.}~\bibnamefont
  {Herr}}, \bibinfo {author} {\bibfnamefont {V.}~\bibnamefont {Brasch}},
  \bibinfo {author} {\bibfnamefont {J.~D.}\ \bibnamefont {Jost}}, \bibinfo
  {author} {\bibfnamefont {C.~Y.}\ \bibnamefont {Wang}}, \bibinfo {author}
  {\bibfnamefont {N.~M.}\ \bibnamefont {Kondratiev}}, \bibinfo {author}
  {\bibfnamefont {M.~L.}\ \bibnamefont {Gorodetsky}}, \ and\ \bibinfo {author}
  {\bibfnamefont {T.~J.}\ \bibnamefont {Kippenberg}},\ }\href {\doibase
  10.1038/nphoton.2013.343} {\bibfield  {journal} {\bibinfo  {journal} {Nature
  Photonics}\ }\textbf {\bibinfo {volume} {8}},\ \bibinfo {pages} {145}
  (\bibinfo {year} {2014})}\BibitemShut {NoStop}%
\bibitem [{\citenamefont {Lugiato}\ and\ \citenamefont
  {Lefever}(1987)}]{LL-PRL-87}%
  \BibitemOpen
  \bibfield  {author} {\bibinfo {author} {\bibfnamefont {L.~A.}\ \bibnamefont
  {Lugiato}}\ and\ \bibinfo {author} {\bibfnamefont {R.}~\bibnamefont
  {Lefever}},\ }\href {\doibase 10.1103/PhysRevLett.58.2209} {\bibfield
  {journal} {\bibinfo  {journal} {Phys. Rev. Lett.}\ }\textbf {\bibinfo
  {volume} {58}},\ \bibinfo {pages} {2209} (\bibinfo {year}
  {1987})}\BibitemShut {NoStop}%
\bibitem [{\citenamefont {Haelterman}\ \emph {et~al.}(1992)\citenamefont
  {Haelterman}, \citenamefont {Trillo},\ and\ \citenamefont
  {Wabnitz}}]{HTW-OC-92}%
  \BibitemOpen
  \bibfield  {author} {\bibinfo {author} {\bibfnamefont {M.}~\bibnamefont
  {Haelterman}}, \bibinfo {author} {\bibfnamefont {S.}~\bibnamefont {Trillo}},
  \ and\ \bibinfo {author} {\bibfnamefont {S.}~\bibnamefont {Wabnitz}},\ }\href
  {\doibase https://doi.org/10.1016/0030-4018(92)90367-Z} {\bibfield  {journal}
  {\bibinfo  {journal} {Optics Communications}\ }\textbf {\bibinfo {volume}
  {91}},\ \bibinfo {pages} {401} (\bibinfo {year} {1992})}\BibitemShut
  {NoStop}%
\bibitem [{\citenamefont {Chembo}\ \emph {et~al.}(2010)\citenamefont {Chembo},
  \citenamefont {Strekalov},\ and\ \citenamefont {Yu}}]{CSY-PRL-10}%
  \BibitemOpen
  \bibfield  {author} {\bibinfo {author} {\bibfnamefont {Y.~K.}\ \bibnamefont
  {Chembo}}, \bibinfo {author} {\bibfnamefont {D.~V.}\ \bibnamefont
  {Strekalov}}, \ and\ \bibinfo {author} {\bibfnamefont {N.}~\bibnamefont
  {Yu}},\ }\href {\doibase 10.1103/PhysRevLett.104.103902} {\bibfield
  {journal} {\bibinfo  {journal} {Phys. Rev. Lett.}\ }\textbf {\bibinfo
  {volume} {104}},\ \bibinfo {pages} {103902} (\bibinfo {year}
  {2010})}\BibitemShut {NoStop}%
\bibitem [{\citenamefont {Chembo}\ and\ \citenamefont {Yu}(2010)}]{CY-PRA-10}%
  \BibitemOpen
  \bibfield  {author} {\bibinfo {author} {\bibfnamefont {Y.~K.}\ \bibnamefont
  {Chembo}}\ and\ \bibinfo {author} {\bibfnamefont {N.}~\bibnamefont {Yu}},\
  }\href {\doibase 10.1103/PhysRevA.82.033801} {\bibfield  {journal} {\bibinfo
  {journal} {Phys. Rev. A}\ }\textbf {\bibinfo {volume} {82}},\ \bibinfo
  {pages} {033801} (\bibinfo {year} {2010})}\BibitemShut {NoStop}%
\bibitem [{\citenamefont {Chembo}\ and\ \citenamefont
  {Menyuk}(2013)}]{CM-PRA-13}%
  \BibitemOpen
  \bibfield  {author} {\bibinfo {author} {\bibfnamefont {Y.~K.}\ \bibnamefont
  {Chembo}}\ and\ \bibinfo {author} {\bibfnamefont {C.~R.}\ \bibnamefont
  {Menyuk}},\ }\href {\doibase 10.1103/PhysRevA.87.053852} {\bibfield
  {journal} {\bibinfo  {journal} {Phys. Rev. A}\ }\textbf {\bibinfo {volume}
  {87}},\ \bibinfo {pages} {053852} (\bibinfo {year} {2013})}\BibitemShut
  {NoStop}%
\bibitem [{\citenamefont {Coen}\ \emph {et~al.}(2013)\citenamefont {Coen},
  \citenamefont {Randle}, \citenamefont {Sylvestre},\ and\ \citenamefont
  {Erkintalo}}]{CRSE-OL-13}%
  \BibitemOpen
  \bibfield  {author} {\bibinfo {author} {\bibfnamefont {S.}~\bibnamefont
  {Coen}}, \bibinfo {author} {\bibfnamefont {H.~G.}\ \bibnamefont {Randle}},
  \bibinfo {author} {\bibfnamefont {T.}~\bibnamefont {Sylvestre}}, \ and\
  \bibinfo {author} {\bibfnamefont {M.}~\bibnamefont {Erkintalo}},\ }\href
  {\doibase 10.1364/OL.38.000037} {\bibfield  {journal} {\bibinfo  {journal}
  {Opt. Lett.}\ }\textbf {\bibinfo {volume} {38}},\ \bibinfo {pages} {37}
  (\bibinfo {year} {2013})}\BibitemShut {NoStop}%
\bibitem [{\citenamefont {Columbo}\ \emph {et~al.}(2021)\citenamefont
  {Columbo}, \citenamefont {Piccardo}, \citenamefont {Prati}, \citenamefont
  {Lugiato}, \citenamefont {Brambilla}, \citenamefont {Gatti}, \citenamefont
  {Silvestri}, \citenamefont {Gioannini}, \citenamefont
  {Opa\ifmmode~\check{c}\else \v{c}\fi{}ak}, \citenamefont {Schwarz},\ and\
  \citenamefont {Capasso}}]{CPP-PRL-21}%
  \BibitemOpen
  \bibfield  {author} {\bibinfo {author} {\bibfnamefont {L.}~\bibnamefont
  {Columbo}}, \bibinfo {author} {\bibfnamefont {M.}~\bibnamefont {Piccardo}},
  \bibinfo {author} {\bibfnamefont {F.}~\bibnamefont {Prati}}, \bibinfo
  {author} {\bibfnamefont {L.~A.}\ \bibnamefont {Lugiato}}, \bibinfo {author}
  {\bibfnamefont {M.}~\bibnamefont {Brambilla}}, \bibinfo {author}
  {\bibfnamefont {A.}~\bibnamefont {Gatti}}, \bibinfo {author} {\bibfnamefont
  {C.}~\bibnamefont {Silvestri}}, \bibinfo {author} {\bibfnamefont
  {M.}~\bibnamefont {Gioannini}}, \bibinfo {author} {\bibfnamefont
  {N.}~\bibnamefont {Opa\ifmmode~\check{c}\else \v{c}\fi{}ak}}, \bibinfo
  {author} {\bibfnamefont {B.}~\bibnamefont {Schwarz}}, \ and\ \bibinfo
  {author} {\bibfnamefont {F.}~\bibnamefont {Capasso}},\ }\href {\doibase
  10.1103/PhysRevLett.126.173903} {\bibfield  {journal} {\bibinfo  {journal}
  {Phys. Rev. Lett.}\ }\textbf {\bibinfo {volume} {126}},\ \bibinfo {pages}
  {173903} (\bibinfo {year} {2021})}\BibitemShut {NoStop}%
\bibitem [{\citenamefont {Lobanov}\ \emph {et~al.}(2015)\citenamefont
  {Lobanov}, \citenamefont {Lihachev}, \citenamefont {Kippenberg},\ and\
  \citenamefont {Gorodetsky}}]{LLK_OE_15}%
  \BibitemOpen
  \bibfield  {author} {\bibinfo {author} {\bibfnamefont {V.}~\bibnamefont
  {Lobanov}}, \bibinfo {author} {\bibfnamefont {G.}~\bibnamefont {Lihachev}},
  \bibinfo {author} {\bibfnamefont {T.~J.}\ \bibnamefont {Kippenberg}}, \ and\
  \bibinfo {author} {\bibfnamefont {M.}~\bibnamefont {Gorodetsky}},\ }\href
  {\doibase 10.1364/OE.23.007713} {\bibfield  {journal} {\bibinfo  {journal}
  {Opt. Express}\ }\textbf {\bibinfo {volume} {23}},\ \bibinfo {pages} {7713}
  (\bibinfo {year} {2015})}\BibitemShut {NoStop}%
\bibitem [{\citenamefont {Xue}\ \emph {et~al.}(2015)\citenamefont {Xue},
  \citenamefont {Xuan}, \citenamefont {Liu}, \citenamefont {Wang},
  \citenamefont {Chen}, \citenamefont {Wang}, \citenamefont {Leaird},
  \citenamefont {Qi},\ and\ \citenamefont {Weiner}}]{XXL_NP_15}%
  \BibitemOpen
  \bibfield  {author} {\bibinfo {author} {\bibfnamefont {X.}~\bibnamefont
  {Xue}}, \bibinfo {author} {\bibfnamefont {Y.}~\bibnamefont {Xuan}}, \bibinfo
  {author} {\bibfnamefont {Y.}~\bibnamefont {Liu}}, \bibinfo {author}
  {\bibfnamefont {P.-H.}\ \bibnamefont {Wang}}, \bibinfo {author}
  {\bibfnamefont {S.}~\bibnamefont {Chen}}, \bibinfo {author} {\bibfnamefont
  {J.}~\bibnamefont {Wang}}, \bibinfo {author} {\bibfnamefont {D.~E.}\
  \bibnamefont {Leaird}}, \bibinfo {author} {\bibfnamefont {M.}~\bibnamefont
  {Qi}}, \ and\ \bibinfo {author} {\bibfnamefont {A.~M.}\ \bibnamefont
  {Weiner}},\ }\href {https://doi.org/10.1038/nphoton.2015.137} {\bibfield
  {journal} {\bibinfo  {journal} {Nature Photonics}\ }\textbf {\bibinfo
  {volume} {9}},\ \bibinfo {pages} {594 EP } (\bibinfo {year} {2015})},\
  \bibinfo {note} {article}\BibitemShut {NoStop}%
\bibitem [{\citenamefont {Parra-Rivas}\ \emph {et~al.}(2016)\citenamefont
  {Parra-Rivas}, \citenamefont {Knobloch}, \citenamefont {Gomila},\ and\
  \citenamefont {Gelens}}]{PKG_PRA_16}%
  \BibitemOpen
  \bibfield  {author} {\bibinfo {author} {\bibfnamefont {P.}~\bibnamefont
  {Parra-Rivas}}, \bibinfo {author} {\bibfnamefont {E.}~\bibnamefont
  {Knobloch}}, \bibinfo {author} {\bibfnamefont {D.}~\bibnamefont {Gomila}}, \
  and\ \bibinfo {author} {\bibfnamefont {L.}~\bibnamefont {Gelens}},\ }\href
  {\doibase 10.1103/PhysRevA.93.063839} {\bibfield  {journal} {\bibinfo
  {journal} {Phys. Rev. A}\ }\textbf {\bibinfo {volume} {93}},\ \bibinfo
  {pages} {063839} (\bibinfo {year} {2016})}\BibitemShut {NoStop}%
\bibitem [{\citenamefont {Garbin}\ \emph {et~al.}(2017)\citenamefont {Garbin},
  \citenamefont {Wang}, \citenamefont {Murdoch}, \citenamefont {Oppo},
  \citenamefont {Coen},\ and\ \citenamefont {Erkintalo}}]{GWM_EPJD_17}%
  \BibitemOpen
  \bibfield  {author} {\bibinfo {author} {\bibfnamefont {B.}~\bibnamefont
  {Garbin}}, \bibinfo {author} {\bibfnamefont {Y.}~\bibnamefont {Wang}},
  \bibinfo {author} {\bibfnamefont {S.~G.}\ \bibnamefont {Murdoch}}, \bibinfo
  {author} {\bibfnamefont {G.-L.}\ \bibnamefont {Oppo}}, \bibinfo {author}
  {\bibfnamefont {S.}~\bibnamefont {Coen}}, \ and\ \bibinfo {author}
  {\bibfnamefont {M.}~\bibnamefont {Erkintalo}},\ }\href {\doibase
  10.1140/epjd/e2017-80133-7} {\bibfield  {journal} {\bibinfo  {journal} {The
  European Physical Journal D}\ }\textbf {\bibinfo {volume} {71}},\ \bibinfo
  {pages} {240} (\bibinfo {year} {2017})}\BibitemShut {NoStop}%
\bibitem [{\citenamefont {Parra-Rivas}\ \emph {et~al.}(2014)\citenamefont
  {Parra-Rivas}, \citenamefont {Gomila}, \citenamefont {Leo}, \citenamefont
  {Coen},\ and\ \citenamefont {Gelens}}]{PGL-OL-14}%
  \BibitemOpen
  \bibfield  {author} {\bibinfo {author} {\bibfnamefont {P.}~\bibnamefont
  {Parra-Rivas}}, \bibinfo {author} {\bibfnamefont {D.}~\bibnamefont {Gomila}},
  \bibinfo {author} {\bibfnamefont {F.}~\bibnamefont {Leo}}, \bibinfo {author}
  {\bibfnamefont {S.}~\bibnamefont {Coen}}, \ and\ \bibinfo {author}
  {\bibfnamefont {L.}~\bibnamefont {Gelens}},\ }\href {\doibase
  10.1364/OL.39.002971} {\bibfield  {journal} {\bibinfo  {journal} {Opt.
  Lett.}\ }\textbf {\bibinfo {volume} {39}},\ \bibinfo {pages} {2971} (\bibinfo
  {year} {2014})}\BibitemShut {NoStop}%
\bibitem [{\citenamefont {Parra-Rivas}\ \emph {et~al.}(2017)\citenamefont
  {Parra-Rivas}, \citenamefont {Gomila},\ and\ \citenamefont
  {Gelens}}]{PGG_PRA_17}%
  \BibitemOpen
  \bibfield  {author} {\bibinfo {author} {\bibfnamefont {P.}~\bibnamefont
  {Parra-Rivas}}, \bibinfo {author} {\bibfnamefont {D.}~\bibnamefont {Gomila}},
  \ and\ \bibinfo {author} {\bibfnamefont {L.}~\bibnamefont {Gelens}},\ }\href
  {\doibase 10.1103/PhysRevA.95.053863} {\bibfield  {journal} {\bibinfo
  {journal} {Phys. Rev. A}\ }\textbf {\bibinfo {volume} {95}},\ \bibinfo
  {pages} {053863} (\bibinfo {year} {2017})}\BibitemShut {NoStop}%
\bibitem [{\citenamefont {Tlidi}\ and\ \citenamefont
  {Gelens}(2010)}]{TG-OL-10}%
  \BibitemOpen
  \bibfield  {author} {\bibinfo {author} {\bibfnamefont {M.}~\bibnamefont
  {Tlidi}}\ and\ \bibinfo {author} {\bibfnamefont {L.}~\bibnamefont {Gelens}},\
  }\href {\doibase 10.1364/OL.35.000306} {\bibfield  {journal} {\bibinfo
  {journal} {Opt. Lett.}\ }\textbf {\bibinfo {volume} {35}},\ \bibinfo {pages}
  {306} (\bibinfo {year} {2010})}\BibitemShut {NoStop}%
\bibitem [{\citenamefont {Schelte}\ \emph
  {et~al.}(2019{\natexlab{a}})\citenamefont {Schelte}, \citenamefont {Pimenov},
  \citenamefont {Vladimirov}, \citenamefont {Javaloyes},\ and\ \citenamefont
  {Gurevich}}]{SPV-OL-19}%
  \BibitemOpen
  \bibfield  {author} {\bibinfo {author} {\bibfnamefont {C.}~\bibnamefont
  {Schelte}}, \bibinfo {author} {\bibfnamefont {A.}~\bibnamefont {Pimenov}},
  \bibinfo {author} {\bibfnamefont {A.~G.}\ \bibnamefont {Vladimirov}},
  \bibinfo {author} {\bibfnamefont {J.}~\bibnamefont {Javaloyes}}, \ and\
  \bibinfo {author} {\bibfnamefont {S.~V.}\ \bibnamefont {Gurevich}},\ }\href
  {\doibase 10.1364/OL.44.004925} {\bibfield  {journal} {\bibinfo  {journal}
  {Opt. Lett.}\ }\textbf {\bibinfo {volume} {44}},\ \bibinfo {pages} {4925}
  (\bibinfo {year} {2019}{\natexlab{a}})}\BibitemShut {NoStop}%
\bibitem [{\citenamefont {Seidel}\ \emph
  {et~al.}(2022{\natexlab{a}})\citenamefont {Seidel}, \citenamefont
  {Javaloyes},\ and\ \citenamefont {Gurevich}}]{SJG-OL-22}%
  \BibitemOpen
  \bibfield  {author} {\bibinfo {author} {\bibfnamefont {T.~G.}\ \bibnamefont
  {Seidel}}, \bibinfo {author} {\bibfnamefont {J.}~\bibnamefont {Javaloyes}}, \
  and\ \bibinfo {author} {\bibfnamefont {S.~V.}\ \bibnamefont {Gurevich}},\
  }\href {\doibase 10.1364/OL.457777} {\bibfield  {journal} {\bibinfo
  {journal} {Opt. Lett.}\ }\textbf {\bibinfo {volume} {47}},\ \bibinfo {pages}
  {2979} (\bibinfo {year} {2022}{\natexlab{a}})}\BibitemShut {NoStop}%
\bibitem [{\citenamefont {Koch}\ \emph {et~al.}(2022)\citenamefont {Koch},
  \citenamefont {Seidel}, \citenamefont {Gurevich},\ and\ \citenamefont
  {Javaloyes}}]{KSG-OL-22}%
  \BibitemOpen
  \bibfield  {author} {\bibinfo {author} {\bibfnamefont {E.}~\bibnamefont
  {Koch}}, \bibinfo {author} {\bibfnamefont {T.}~\bibnamefont {Seidel}},
  \bibinfo {author} {\bibfnamefont {S.}~\bibnamefont {Gurevich}}, \ and\
  \bibinfo {author} {\bibfnamefont {J.}~\bibnamefont {Javaloyes}},\ }\href
  {\doibase 10.1364/OL.468236} {\bibfield  {journal} {\bibinfo  {journal}
  {Optics Letters}\ }\textbf {\bibinfo {volume} {47}},\ \bibinfo {pages} {4343}
  (\bibinfo {year} {2022})}\BibitemShut {NoStop}%
\bibitem [{\citenamefont {Koch}\ \emph {et~al.}(2023)\citenamefont {Koch},
  \citenamefont {Seidel}, \citenamefont {Javaloyes},\ and\ \citenamefont
  {Gurevich}}]{KSJ-CHA-23}%
  \BibitemOpen
  \bibfield  {author} {\bibinfo {author} {\bibfnamefont {E.}~\bibnamefont
  {Koch}}, \bibinfo {author} {\bibfnamefont {T.}~\bibnamefont {Seidel}},
  \bibinfo {author} {\bibfnamefont {J.}~\bibnamefont {Javaloyes}}, \ and\
  \bibinfo {author} {\bibfnamefont {S.}~\bibnamefont {Gurevich}},\ }\href
  {\doibase 10.1063/5.0143562} {\bibfield  {journal} {\bibinfo  {journal}
  {Chaos: An Interdisciplinary Journal of Nonlinear Science}\ }\textbf
  {\bibinfo {volume} {33}} (\bibinfo {year} {2023}),\
  10.1063/5.0143562}\BibitemShut {NoStop}%
\bibitem [{\citenamefont {Koch}\ \emph {et~al.}(2026)\citenamefont {Koch},
  \citenamefont {Greve}, \citenamefont {Javaloyes},\ and\ \citenamefont
  {Gurevich}}]{KGJ-CSF-26}%
  \BibitemOpen
  \bibfield  {author} {\bibinfo {author} {\bibfnamefont {E.~R.}\ \bibnamefont
  {Koch}}, \bibinfo {author} {\bibfnamefont {D.}~\bibnamefont {Greve}},
  \bibinfo {author} {\bibfnamefont {J.}~\bibnamefont {Javaloyes}}, \ and\
  \bibinfo {author} {\bibfnamefont {S.~V.}\ \bibnamefont {Gurevich}},\ }\href
  {\doibase https://doi.org/10.1016/j.chaos.2025.117595} {\bibfield  {journal}
  {\bibinfo  {journal} {Chaos, Solitons \& Fractals}\ }\textbf {\bibinfo
  {volume} {202}},\ \bibinfo {pages} {117595} (\bibinfo {year}
  {2026})}\BibitemShut {NoStop}%
\bibitem [{\citenamefont {Seidel}\ \emph
  {et~al.}(2022{\natexlab{b}})\citenamefont {Seidel}, \citenamefont
  {Gurevich},\ and\ \citenamefont {Javaloyes}}]{SGJ-PRL-22}%
  \BibitemOpen
  \bibfield  {author} {\bibinfo {author} {\bibfnamefont {T.~G.}\ \bibnamefont
  {Seidel}}, \bibinfo {author} {\bibfnamefont {S.~V.}\ \bibnamefont
  {Gurevich}}, \ and\ \bibinfo {author} {\bibfnamefont {J.}~\bibnamefont
  {Javaloyes}},\ }\href {\doibase 10.1103/PhysRevLett.128.083901} {\bibfield
  {journal} {\bibinfo  {journal} {Phys. Rev. Lett.}\ }\textbf {\bibinfo
  {volume} {128}},\ \bibinfo {pages} {083901} (\bibinfo {year}
  {2022}{\natexlab{b}})}\BibitemShut {NoStop}%
\bibitem [{\citenamefont {Seidel}\ \emph {et~al.}(2024)\citenamefont {Seidel},
  \citenamefont {Javaloyes},\ and\ \citenamefont {Gurevich}}]{SJG-OL-24}%
  \BibitemOpen
  \bibfield  {author} {\bibinfo {author} {\bibfnamefont {T.~G.}\ \bibnamefont
  {Seidel}}, \bibinfo {author} {\bibfnamefont {J.}~\bibnamefont {Javaloyes}}, \
  and\ \bibinfo {author} {\bibfnamefont {S.~V.}\ \bibnamefont {Gurevich}},\
  }\href {\doibase 10.1364/OL.538135} {\bibfield  {journal} {\bibinfo
  {journal} {Opt. Lett.}\ }\textbf {\bibinfo {volume} {49}},\ \bibinfo {pages}
  {7008} (\bibinfo {year} {2024})}\BibitemShut {NoStop}%
\bibitem [{\citenamefont {Odent}\ \emph {et~al.}(2014)\citenamefont {Odent},
  \citenamefont {Tlidi}, \citenamefont {Clerc}, \citenamefont {Glorieux},\ and\
  \citenamefont {Louvergneaux}}]{OTC-PRA-14}%
  \BibitemOpen
  \bibfield  {author} {\bibinfo {author} {\bibfnamefont {V.}~\bibnamefont
  {Odent}}, \bibinfo {author} {\bibfnamefont {M.}~\bibnamefont {Tlidi}},
  \bibinfo {author} {\bibfnamefont {M.~G.}\ \bibnamefont {Clerc}}, \bibinfo
  {author} {\bibfnamefont {P.}~\bibnamefont {Glorieux}}, \ and\ \bibinfo
  {author} {\bibfnamefont {E.}~\bibnamefont {Louvergneaux}},\ }\href {\doibase
  10.1103/PhysRevA.90.011806} {\bibfield  {journal} {\bibinfo  {journal} {Phys.
  Rev. A}\ }\textbf {\bibinfo {volume} {90}},\ \bibinfo {pages} {011806}
  (\bibinfo {year} {2014})}\BibitemShut {NoStop}%
\bibitem [{\citenamefont {Jang}\ \emph {et~al.}(2015)\citenamefont {Jang},
  \citenamefont {Erkintalo}, \citenamefont {Coen},\ and\ \citenamefont
  {Murdoch}}]{JEC-NAC-15}%
  \BibitemOpen
  \bibfield  {author} {\bibinfo {author} {\bibfnamefont {J.~K.}\ \bibnamefont
  {Jang}}, \bibinfo {author} {\bibfnamefont {M.}~\bibnamefont {Erkintalo}},
  \bibinfo {author} {\bibfnamefont {S.}~\bibnamefont {Coen}}, \ and\ \bibinfo
  {author} {\bibfnamefont {S.~G.}\ \bibnamefont {Murdoch}},\ }\href
  {http://dx.doi.org/10.1038/ncomms8370} {\bibfield  {journal} {\bibinfo
  {journal} {Nat Commun}\ }\textbf {\bibinfo {volume} {6}} (\bibinfo {year}
  {2015})},\ \bibinfo {note} {article}\BibitemShut {NoStop}%
\bibitem [{\citenamefont {Cole}\ \emph {et~al.}(2018)\citenamefont {Cole},
  \citenamefont {Stone}, \citenamefont {Erkintalo}, \citenamefont {Yang},
  \citenamefont {Yi}, \citenamefont {Vahala},\ and\ \citenamefont
  {Papp}}]{CSE-Optica-18}%
  \BibitemOpen
  \bibfield  {author} {\bibinfo {author} {\bibfnamefont {D.~C.}\ \bibnamefont
  {Cole}}, \bibinfo {author} {\bibfnamefont {J.~R.}\ \bibnamefont {Stone}},
  \bibinfo {author} {\bibfnamefont {M.}~\bibnamefont {Erkintalo}}, \bibinfo
  {author} {\bibfnamefont {K.~Y.}\ \bibnamefont {Yang}}, \bibinfo {author}
  {\bibfnamefont {X.}~\bibnamefont {Yi}}, \bibinfo {author} {\bibfnamefont
  {K.~J.}\ \bibnamefont {Vahala}}, \ and\ \bibinfo {author} {\bibfnamefont
  {S.~B.}\ \bibnamefont {Papp}},\ }\href {\doibase 10.1364/OPTICA.5.001304}
  {\bibfield  {journal} {\bibinfo  {journal} {Optica}\ }\textbf {\bibinfo
  {volume} {5}},\ \bibinfo {pages} {1304} (\bibinfo {year} {2018})}\BibitemShut
  {NoStop}%
\bibitem [{\citenamefont {Hendry}\ \emph {et~al.}(2018)\citenamefont {Hendry},
  \citenamefont {Chen}, \citenamefont {Wang}, \citenamefont {Garbin},
  \citenamefont {Javaloyes}, \citenamefont {Oppo}, \citenamefont {Coen},
  \citenamefont {Murdoch},\ and\ \citenamefont {Erkintalo}}]{Hendry2018}%
  \BibitemOpen
  \bibfield  {author} {\bibinfo {author} {\bibfnamefont {I.}~\bibnamefont
  {Hendry}}, \bibinfo {author} {\bibfnamefont {W.}~\bibnamefont {Chen}},
  \bibinfo {author} {\bibfnamefont {Y.}~\bibnamefont {Wang}}, \bibinfo {author}
  {\bibfnamefont {B.}~\bibnamefont {Garbin}}, \bibinfo {author} {\bibfnamefont
  {J.}~\bibnamefont {Javaloyes}}, \bibinfo {author} {\bibfnamefont {G.-L.}\
  \bibnamefont {Oppo}}, \bibinfo {author} {\bibfnamefont {S.}~\bibnamefont
  {Coen}}, \bibinfo {author} {\bibfnamefont {S.~G.}\ \bibnamefont {Murdoch}}, \
  and\ \bibinfo {author} {\bibfnamefont {M.}~\bibnamefont {Erkintalo}},\ }\href
  {\doibase 10.1103/PhysRevA.97.053834} {\bibfield  {journal} {\bibinfo
  {journal} {Phys. Rev. A}\ }\textbf {\bibinfo {volume} {97}},\ \bibinfo
  {pages} {053834} (\bibinfo {year} {2018})}\BibitemShut {NoStop}%
\bibitem [{\citenamefont {Tabbert}\ \emph {et~al.}(2019)\citenamefont
  {Tabbert}, \citenamefont {Frohoff-H\"ulsmann}, \citenamefont {Panajotov},
  \citenamefont {Tlidi},\ and\ \citenamefont {Gurevich}}]{TFHP-PRA-19}%
  \BibitemOpen
  \bibfield  {author} {\bibinfo {author} {\bibfnamefont {F.}~\bibnamefont
  {Tabbert}}, \bibinfo {author} {\bibfnamefont {T.}~\bibnamefont
  {Frohoff-H\"ulsmann}}, \bibinfo {author} {\bibfnamefont {K.}~\bibnamefont
  {Panajotov}}, \bibinfo {author} {\bibfnamefont {M.}~\bibnamefont {Tlidi}}, \
  and\ \bibinfo {author} {\bibfnamefont {S.~V.}\ \bibnamefont {Gurevich}},\
  }\href {\doibase 10.1103/PhysRevA.100.013818} {\bibfield  {journal} {\bibinfo
   {journal} {Phys. Rev. A}\ }\textbf {\bibinfo {volume} {100}},\ \bibinfo
  {pages} {013818} (\bibinfo {year} {2019})}\BibitemShut {NoStop}%
\bibitem [{\citenamefont {Dolinina}\ \emph {et~al.}(2024)\citenamefont
  {Dolinina}, \citenamefont {Huyet}, \citenamefont {Turaev},\ and\
  \citenamefont {Vladimirov}}]{DHT-OL-24}%
  \BibitemOpen
  \bibfield  {author} {\bibinfo {author} {\bibfnamefont {D.~A.}\ \bibnamefont
  {Dolinina}}, \bibinfo {author} {\bibfnamefont {G.}~\bibnamefont {Huyet}},
  \bibinfo {author} {\bibfnamefont {D.}~\bibnamefont {Turaev}}, \ and\ \bibinfo
  {author} {\bibfnamefont {A.~G.}\ \bibnamefont {Vladimirov}},\ }\href
  {\doibase 10.1364/OL.529083} {\bibfield  {journal} {\bibinfo  {journal} {Opt.
  Lett.}\ }\textbf {\bibinfo {volume} {49}},\ \bibinfo {pages} {4050} (\bibinfo
  {year} {2024})}\BibitemShut {NoStop}%
\bibitem [{\citenamefont {Obrzud}\ \emph {et~al.}(2017)\citenamefont {Obrzud},
  \citenamefont {Lecomte},\ and\ \citenamefont {Herr}}]{OLH-NatPhot-17}%
  \BibitemOpen
  \bibfield  {author} {\bibinfo {author} {\bibfnamefont {E.}~\bibnamefont
  {Obrzud}}, \bibinfo {author} {\bibfnamefont {S.}~\bibnamefont {Lecomte}}, \
  and\ \bibinfo {author} {\bibfnamefont {T.}~\bibnamefont {Herr}},\ }\href
  {\doibase 10.1038/nphoton.2017.140} {\bibfield  {journal} {\bibinfo
  {journal} {Nature Photonics}\ }\textbf {\bibinfo {volume} {11}},\ \bibinfo
  {pages} {600} (\bibinfo {year} {2017})}\BibitemShut {NoStop}%
\bibitem [{\citenamefont {Ivars}\ \emph {et~al.}(2024)\citenamefont {Ivars},
  \citenamefont {Mili\'an}, \citenamefont {Botey}, \citenamefont {Herrero},\
  and\ \citenamefont {Staliunas}}]{IMB-PRL-24}%
  \BibitemOpen
  \bibfield  {author} {\bibinfo {author} {\bibfnamefont {S.~B.}\ \bibnamefont
  {Ivars}}, \bibinfo {author} {\bibfnamefont {C.}~\bibnamefont {Mili\'an}},
  \bibinfo {author} {\bibfnamefont {M.}~\bibnamefont {Botey}}, \bibinfo
  {author} {\bibfnamefont {R.}~\bibnamefont {Herrero}}, \ and\ \bibinfo
  {author} {\bibfnamefont {K.}~\bibnamefont {Staliunas}},\ }\href {\doibase
  10.1103/PhysRevLett.133.093802} {\bibfield  {journal} {\bibinfo  {journal}
  {Phys. Rev. Lett.}\ }\textbf {\bibinfo {volume} {133}},\ \bibinfo {pages}
  {093802} (\bibinfo {year} {2024})}\BibitemShut {NoStop}%
\bibitem [{\citenamefont {Bersch}\ \emph {et~al.}(2012)\citenamefont {Bersch},
  \citenamefont {Onishchukov},\ and\ \citenamefont {Peschel}}]{BOP-PRL-12}%
  \BibitemOpen
  \bibfield  {author} {\bibinfo {author} {\bibfnamefont {C.}~\bibnamefont
  {Bersch}}, \bibinfo {author} {\bibfnamefont {G.}~\bibnamefont {Onishchukov}},
  \ and\ \bibinfo {author} {\bibfnamefont {U.}~\bibnamefont {Peschel}},\ }\href
  {\doibase 10.1103/PhysRevLett.109.093903} {\bibfield  {journal} {\bibinfo
  {journal} {Phys. Rev. Lett.}\ }\textbf {\bibinfo {volume} {109}},\ \bibinfo
  {pages} {093903} (\bibinfo {year} {2012})}\BibitemShut {NoStop}%
\bibitem [{\citenamefont {Tusnin}\ \emph {et~al.}(2020)\citenamefont {Tusnin},
  \citenamefont {Tikan},\ and\ \citenamefont {Kippenberg}}]{TTK-PRA-20}%
  \BibitemOpen
  \bibfield  {author} {\bibinfo {author} {\bibfnamefont {A.~K.}\ \bibnamefont
  {Tusnin}}, \bibinfo {author} {\bibfnamefont {A.~M.}\ \bibnamefont {Tikan}}, \
  and\ \bibinfo {author} {\bibfnamefont {T.~J.}\ \bibnamefont {Kippenberg}},\
  }\href {\doibase 10.1103/PhysRevA.102.023518} {\bibfield  {journal} {\bibinfo
   {journal} {Phys. Rev. A}\ }\textbf {\bibinfo {volume} {102}},\ \bibinfo
  {pages} {023518} (\bibinfo {year} {2020})}\BibitemShut {NoStop}%
\bibitem [{\citenamefont {Sun}\ \emph {et~al.}(2022)\citenamefont {Sun},
  \citenamefont {Parra-Rivas}, \citenamefont {Ferraro}, \citenamefont
  {Mangini}, \citenamefont {Zitelli}, \citenamefont {Jauberteau}, \citenamefont
  {Talenti},\ and\ \citenamefont {Wabnitz}}]{SPRF-OL-22}%
  \BibitemOpen
  \bibfield  {author} {\bibinfo {author} {\bibfnamefont {Y.}~\bibnamefont
  {Sun}}, \bibinfo {author} {\bibfnamefont {P.}~\bibnamefont {Parra-Rivas}},
  \bibinfo {author} {\bibfnamefont {M.}~\bibnamefont {Ferraro}}, \bibinfo
  {author} {\bibfnamefont {F.}~\bibnamefont {Mangini}}, \bibinfo {author}
  {\bibfnamefont {M.}~\bibnamefont {Zitelli}}, \bibinfo {author} {\bibfnamefont
  {R.}~\bibnamefont {Jauberteau}}, \bibinfo {author} {\bibfnamefont {F.~R.}\
  \bibnamefont {Talenti}}, \ and\ \bibinfo {author} {\bibfnamefont
  {S.}~\bibnamefont {Wabnitz}},\ }\href {\doibase 10.1364/OL.472900} {\bibfield
   {journal} {\bibinfo  {journal} {Opt. Lett.}\ }\textbf {\bibinfo {volume}
  {47}},\ \bibinfo {pages} {6353} (\bibinfo {year} {2022})}\BibitemShut
  {NoStop}%
\bibitem [{\citenamefont {Sun}\ \emph {et~al.}(2023{\natexlab{a}})\citenamefont
  {Sun}, \citenamefont {Wabnitz},\ and\ \citenamefont
  {Parra-Rivas}}]{SWPR-OL-23}%
  \BibitemOpen
  \bibfield  {author} {\bibinfo {author} {\bibfnamefont {Y.}~\bibnamefont
  {Sun}}, \bibinfo {author} {\bibfnamefont {S.}~\bibnamefont {Wabnitz}}, \ and\
  \bibinfo {author} {\bibfnamefont {P.}~\bibnamefont {Parra-Rivas}},\ }\href
  {\doibase 10.1364/OL.499907} {\bibfield  {journal} {\bibinfo  {journal} {Opt.
  Lett.}\ }\textbf {\bibinfo {volume} {48}},\ \bibinfo {pages} {5403} (\bibinfo
  {year} {2023}{\natexlab{a}})}\BibitemShut {NoStop}%
\bibitem [{\citenamefont {Sun}\ \emph {et~al.}(2023{\natexlab{b}})\citenamefont
  {Sun}, \citenamefont {Parra-Rivas}, \citenamefont {Ferraro}, \citenamefont
  {Mangini},\ and\ \citenamefont {Wabnitz}}]{SPRF-CSF-23}%
  \BibitemOpen
  \bibfield  {author} {\bibinfo {author} {\bibfnamefont {Y.}~\bibnamefont
  {Sun}}, \bibinfo {author} {\bibfnamefont {P.}~\bibnamefont {Parra-Rivas}},
  \bibinfo {author} {\bibfnamefont {M.}~\bibnamefont {Ferraro}}, \bibinfo
  {author} {\bibfnamefont {F.}~\bibnamefont {Mangini}}, \ and\ \bibinfo
  {author} {\bibfnamefont {S.}~\bibnamefont {Wabnitz}},\ }\href {\doibase
  https://doi.org/10.1016/j.chaos.2023.114064} {\bibfield  {journal} {\bibinfo
  {journal} {Chaos, Solitons \& Fractals}\ }\textbf {\bibinfo {volume} {176}},\
  \bibinfo {pages} {114064} (\bibinfo {year} {2023}{\natexlab{b}})}\BibitemShut
  {NoStop}%
\bibitem [{\citenamefont {Gires}\ and\ \citenamefont
  {Tournois}(1964)}]{GT-CRA-64}%
  \BibitemOpen
  \bibfield  {author} {\bibinfo {author} {\bibfnamefont {F.}~\bibnamefont
  {Gires}}\ and\ \bibinfo {author} {\bibfnamefont {P.}~\bibnamefont
  {Tournois}},\ }\href@noop {} {\bibfield  {journal} {\bibinfo  {journal} {C.
  R. Acad. Sci. Paris}\ ,\ \bibinfo {pages} {6112}} (\bibinfo {year}
  {1964})}\BibitemShut {NoStop}%
\bibitem [{\citenamefont {Schelte}\ \emph
  {et~al.}(2019{\natexlab{b}})\citenamefont {Schelte}, \citenamefont {Camelin},
  \citenamefont {Marconi}, \citenamefont {Garnache}, \citenamefont {Huyet},
  \citenamefont {Beaudoin}, \citenamefont {Sagnes}, \citenamefont {Giudici},
  \citenamefont {Javaloyes},\ and\ \citenamefont {Gurevich}}]{SCM-PRL-19}%
  \BibitemOpen
  \bibfield  {author} {\bibinfo {author} {\bibfnamefont {C.}~\bibnamefont
  {Schelte}}, \bibinfo {author} {\bibfnamefont {P.}~\bibnamefont {Camelin}},
  \bibinfo {author} {\bibfnamefont {M.}~\bibnamefont {Marconi}}, \bibinfo
  {author} {\bibfnamefont {A.}~\bibnamefont {Garnache}}, \bibinfo {author}
  {\bibfnamefont {G.}~\bibnamefont {Huyet}}, \bibinfo {author} {\bibfnamefont
  {G.}~\bibnamefont {Beaudoin}}, \bibinfo {author} {\bibfnamefont
  {I.}~\bibnamefont {Sagnes}}, \bibinfo {author} {\bibfnamefont
  {M.}~\bibnamefont {Giudici}}, \bibinfo {author} {\bibfnamefont
  {J.}~\bibnamefont {Javaloyes}}, \ and\ \bibinfo {author} {\bibfnamefont
  {S.~V.}\ \bibnamefont {Gurevich}},\ }\href {\doibase
  10.1103/PhysRevLett.123.043902} {\bibfield  {journal} {\bibinfo  {journal}
  {Phys. Rev. Lett.}\ }\textbf {\bibinfo {volume} {123}},\ \bibinfo {pages}
  {043902} (\bibinfo {year} {2019}{\natexlab{b}})}\BibitemShut {NoStop}%
\bibitem [{\citenamefont {Vladimirov}\ and\ \citenamefont
  {Dolinina}(2024)}]{VD-PRE-24}%
  \BibitemOpen
  \bibfield  {author} {\bibinfo {author} {\bibfnamefont {A.~G.}\ \bibnamefont
  {Vladimirov}}\ and\ \bibinfo {author} {\bibfnamefont {D.~A.}\ \bibnamefont
  {Dolinina}},\ }\href {\doibase 10.1103/PhysRevE.109.024206} {\bibfield
  {journal} {\bibinfo  {journal} {Phys. Rev. E}\ }\textbf {\bibinfo {volume}
  {109}},\ \bibinfo {pages} {024206} (\bibinfo {year} {2024})}\BibitemShut
  {NoStop}%
\bibitem [{\citenamefont {Mulet}\ and\ \citenamefont
  {Balle}(2005)}]{MB-JQE-05}%
  \BibitemOpen
  \bibfield  {author} {\bibinfo {author} {\bibfnamefont {J.}~\bibnamefont
  {Mulet}}\ and\ \bibinfo {author} {\bibfnamefont {S.}~\bibnamefont {Balle}},\
  }\href {\doibase 10.1109/JQE.2005.853355} {\bibfield  {journal} {\bibinfo
  {journal} {Quantum Electronics, IEEE Journal of}\ }\textbf {\bibinfo {volume}
  {41}},\ \bibinfo {pages} {1148} (\bibinfo {year} {2005})}\BibitemShut
  {NoStop}%
\bibitem [{\citenamefont {Camelin}\ \emph {et~al.}(2018)\citenamefont
  {Camelin}, \citenamefont {Schelte}, \citenamefont {Verschelde}, \citenamefont
  {Garnache}, \citenamefont {Beaudoin}, \citenamefont {Sagnes}, \citenamefont
  {Huyet}, \citenamefont {Javaloyes}, \citenamefont {Gurevich},\ and\
  \citenamefont {Giudici}}]{CSV-OL-18}%
  \BibitemOpen
  \bibfield  {author} {\bibinfo {author} {\bibfnamefont {P.}~\bibnamefont
  {Camelin}}, \bibinfo {author} {\bibfnamefont {C.}~\bibnamefont {Schelte}},
  \bibinfo {author} {\bibfnamefont {A.}~\bibnamefont {Verschelde}}, \bibinfo
  {author} {\bibfnamefont {A.}~\bibnamefont {Garnache}}, \bibinfo {author}
  {\bibfnamefont {G.}~\bibnamefont {Beaudoin}}, \bibinfo {author}
  {\bibfnamefont {I.}~\bibnamefont {Sagnes}}, \bibinfo {author} {\bibfnamefont
  {G.}~\bibnamefont {Huyet}}, \bibinfo {author} {\bibfnamefont
  {J.}~\bibnamefont {Javaloyes}}, \bibinfo {author} {\bibfnamefont {S.~V.}\
  \bibnamefont {Gurevich}}, \ and\ \bibinfo {author} {\bibfnamefont
  {M.}~\bibnamefont {Giudici}},\ }\href {\doibase 10.1364/OL.43.005367}
  {\bibfield  {journal} {\bibinfo  {journal} {Opt. Lett.}\ }\textbf {\bibinfo
  {volume} {43}},\ \bibinfo {pages} {5367} (\bibinfo {year}
  {2018})}\BibitemShut {NoStop}%
\bibitem [{\citenamefont {Schelte}\ \emph {et~al.}(2020)\citenamefont
  {Schelte}, \citenamefont {Hessel}, \citenamefont {Javaloyes},\ and\
  \citenamefont {Gurevich}}]{SHJ-PRAp-20}%
  \BibitemOpen
  \bibfield  {author} {\bibinfo {author} {\bibfnamefont {C.}~\bibnamefont
  {Schelte}}, \bibinfo {author} {\bibfnamefont {D.}~\bibnamefont {Hessel}},
  \bibinfo {author} {\bibfnamefont {J.}~\bibnamefont {Javaloyes}}, \ and\
  \bibinfo {author} {\bibfnamefont {S.~V.}\ \bibnamefont {Gurevich}},\ }\href
  {\doibase 10.1103/PhysRevApplied.13.054050} {\bibfield  {journal} {\bibinfo
  {journal} {Phys. Rev. Applied}\ }\textbf {\bibinfo {volume} {13}},\ \bibinfo
  {pages} {054050} (\bibinfo {year} {2020})}\BibitemShut {NoStop}%
\bibitem [{\citenamefont {Hessel}\ \emph {et~al.}(2021)\citenamefont {Hessel},
  \citenamefont {Gurevich},\ and\ \citenamefont {Javaloyes}}]{HGJ-OL-21}%
  \BibitemOpen
  \bibfield  {author} {\bibinfo {author} {\bibfnamefont {D.}~\bibnamefont
  {Hessel}}, \bibinfo {author} {\bibfnamefont {S.~V.}\ \bibnamefont
  {Gurevich}}, \ and\ \bibinfo {author} {\bibfnamefont {J.}~\bibnamefont
  {Javaloyes}},\ }\href {\doibase 10.1364/OL.428182} {\bibfield  {journal}
  {\bibinfo  {journal} {Opt. Lett.}\ }\textbf {\bibinfo {volume} {46}},\
  \bibinfo {pages} {2557} (\bibinfo {year} {2021})}\BibitemShut {NoStop}%
\bibitem [{\citenamefont {Pikovsky}\ \emph {et~al.}(2001)\citenamefont
  {Pikovsky}, \citenamefont {Rosenblum},\ and\ \citenamefont
  {Kurths}}]{synchrobook}%
  \BibitemOpen
  \bibfield  {author} {\bibinfo {author} {\bibfnamefont {A.}~\bibnamefont
  {Pikovsky}}, \bibinfo {author} {\bibfnamefont {M.}~\bibnamefont {Rosenblum}},
  \ and\ \bibinfo {author} {\bibfnamefont {J.}~\bibnamefont {Kurths}},\
  }\href@noop {} {\emph {\bibinfo {title} {Synchronization : a universal
  concept in nonlinear sciences}}}\ (\bibinfo  {publisher} {Cambridge
  University Press},\ \bibinfo {address} {New York},\ \bibinfo {year}
  {2001})\BibitemShut {NoStop}%
\bibitem [{\citenamefont {Arecchi}\ \emph {et~al.}(1992)\citenamefont
  {Arecchi}, \citenamefont {Giacomelli}, \citenamefont {Lapucci},\ and\
  \citenamefont {Meucci}}]{AGL-PRA-92}%
  \BibitemOpen
  \bibfield  {author} {\bibinfo {author} {\bibfnamefont {F.~T.}\ \bibnamefont
  {Arecchi}}, \bibinfo {author} {\bibfnamefont {G.}~\bibnamefont {Giacomelli}},
  \bibinfo {author} {\bibfnamefont {A.}~\bibnamefont {Lapucci}}, \ and\
  \bibinfo {author} {\bibfnamefont {R.}~\bibnamefont {Meucci}},\ }\href
  {\doibase 10.1103/PhysRevA.45.R4225} {\bibfield  {journal} {\bibinfo
  {journal} {Phys. Rev. A}\ }\textbf {\bibinfo {volume} {45}},\ \bibinfo
  {pages} {R4225} (\bibinfo {year} {1992})}\BibitemShut {NoStop}%
\bibitem [{\citenamefont {Giacomelli}\ and\ \citenamefont
  {Politi}(1996)}]{GP-PRL-96}%
  \BibitemOpen
  \bibfield  {author} {\bibinfo {author} {\bibfnamefont {G.}~\bibnamefont
  {Giacomelli}}\ and\ \bibinfo {author} {\bibfnamefont {A.}~\bibnamefont
  {Politi}},\ }\href {\doibase 10.1103/PhysRevLett.76.2686} {\bibfield
  {journal} {\bibinfo  {journal} {Phys. Rev. Lett.}\ }\textbf {\bibinfo
  {volume} {76}},\ \bibinfo {pages} {2686} (\bibinfo {year}
  {1996})}\BibitemShut {NoStop}%
\bibitem [{\citenamefont {Marino}\ and\ \citenamefont
  {Giacomelli}(2020)}]{FG-PRE-20}%
  \BibitemOpen
  \bibfield  {author} {\bibinfo {author} {\bibfnamefont {F.}~\bibnamefont
  {Marino}}\ and\ \bibinfo {author} {\bibfnamefont {G.}~\bibnamefont
  {Giacomelli}},\ }\href {\doibase 10.1103/PhysRevE.102.052217} {\bibfield
  {journal} {\bibinfo  {journal} {Phys. Rev. E}\ }\textbf {\bibinfo {volume}
  {102}},\ \bibinfo {pages} {052217} (\bibinfo {year} {2020})}\BibitemShut
  {NoStop}%
\bibitem [{\citenamefont {Engelborghs}\ \emph {et~al.}(2002)\citenamefont
  {Engelborghs}, \citenamefont {Luzyanina},\ and\ \citenamefont
  {Roose}}]{DDEBT}%
  \BibitemOpen
  \bibfield  {author} {\bibinfo {author} {\bibfnamefont {K.}~\bibnamefont
  {Engelborghs}}, \bibinfo {author} {\bibfnamefont {T.}~\bibnamefont
  {Luzyanina}}, \ and\ \bibinfo {author} {\bibfnamefont {D.}~\bibnamefont
  {Roose}},\ }\href {\doibase 10.1145/513001.513002} {\bibfield  {journal}
  {\bibinfo  {journal} {ACM Trans. Math. Softw.}\ }\textbf {\bibinfo {volume}
  {28}},\ \bibinfo {pages} {1} (\bibinfo {year} {2002})}\BibitemShut {NoStop}%
\bibitem [{\citenamefont {Gurevich}\ \emph {et~al.}(2024)\citenamefont
  {Gurevich}, \citenamefont {Maucher},\ and\ \citenamefont
  {Javaloyes}}]{GMJ-PRR-24}%
  \BibitemOpen
  \bibfield  {author} {\bibinfo {author} {\bibfnamefont {S.~V.}\ \bibnamefont
  {Gurevich}}, \bibinfo {author} {\bibfnamefont {F.}~\bibnamefont {Maucher}}, \
  and\ \bibinfo {author} {\bibfnamefont {J.}~\bibnamefont {Javaloyes}},\ }\href
  {\doibase 10.1103/PhysRevResearch.6.013166} {\bibfield  {journal} {\bibinfo
  {journal} {Phys. Rev. Res.}\ }\textbf {\bibinfo {volume} {6}},\ \bibinfo
  {pages} {013166} (\bibinfo {year} {2024})}\BibitemShut {NoStop}%
\bibitem [{\citenamefont {Bai}\ and\ \citenamefont {Breen}(2008)}]{BB-JGT-08}%
  \BibitemOpen
  \bibfield  {author} {\bibinfo {author} {\bibfnamefont {L.}~\bibnamefont
  {Bai}}\ and\ \bibinfo {author} {\bibfnamefont {D.}~\bibnamefont {Breen}},\
  }\href {\doibase 10.1080/2151237x.2008.10129266} {\bibfield  {journal}
  {\bibinfo  {journal} {Journal of Graphics Tools}\ }\textbf {\bibinfo {volume}
  {13}},\ \bibinfo {pages} {53–60} (\bibinfo {year} {2008})}\BibitemShut
  {NoStop}%
\bibitem [{\citenamefont {Haudin}\ \emph {et~al.}(2010)\citenamefont {Haudin},
  \citenamefont {El\'{\i}as}, \citenamefont {Rojas}, \citenamefont
  {Bortolozzo}, \citenamefont {Clerc},\ and\ \citenamefont
  {Residori}}]{HER-PRE-10}%
  \BibitemOpen
  \bibfield  {author} {\bibinfo {author} {\bibfnamefont {F.}~\bibnamefont
  {Haudin}}, \bibinfo {author} {\bibfnamefont {R.~G.}\ \bibnamefont
  {El\'{\i}as}}, \bibinfo {author} {\bibfnamefont {R.~G.}\ \bibnamefont
  {Rojas}}, \bibinfo {author} {\bibfnamefont {U.}~\bibnamefont {Bortolozzo}},
  \bibinfo {author} {\bibfnamefont {M.~G.}\ \bibnamefont {Clerc}}, \ and\
  \bibinfo {author} {\bibfnamefont {S.}~\bibnamefont {Residori}},\ }\href
  {\doibase 10.1103/PhysRevE.81.056203} {\bibfield  {journal} {\bibinfo
  {journal} {Phys. Rev. E}\ }\textbf {\bibinfo {volume} {81}},\ \bibinfo
  {pages} {056203} (\bibinfo {year} {2010})}\BibitemShut {NoStop}%
\bibitem [{\citenamefont {Tabbert}\ \emph {et~al.}(2017)\citenamefont
  {Tabbert}, \citenamefont {Schelte}, \citenamefont {Tlidi},\ and\
  \citenamefont {Gurevich}}]{TST-PRE-17}%
  \BibitemOpen
  \bibfield  {author} {\bibinfo {author} {\bibfnamefont {F.}~\bibnamefont
  {Tabbert}}, \bibinfo {author} {\bibfnamefont {C.}~\bibnamefont {Schelte}},
  \bibinfo {author} {\bibfnamefont {M.}~\bibnamefont {Tlidi}}, \ and\ \bibinfo
  {author} {\bibfnamefont {S.~V.}\ \bibnamefont {Gurevich}},\ }\href {\doibase
  10.1103/PhysRevE.95.032213} {\bibfield  {journal} {\bibinfo  {journal} {Phys.
  Rev. E}\ }\textbf {\bibinfo {volume} {95}},\ \bibinfo {pages} {032213}
  (\bibinfo {year} {2017})}\BibitemShut {NoStop}%
\bibitem [{\citenamefont {Munsberg}\ \emph {et~al.}(2020)\citenamefont
  {Munsberg}, \citenamefont {Javaloyes},\ and\ \citenamefont
  {Gurevich}}]{MJG-Chaos-20}%
  \BibitemOpen
  \bibfield  {author} {\bibinfo {author} {\bibfnamefont {L.}~\bibnamefont
  {Munsberg}}, \bibinfo {author} {\bibfnamefont {J.}~\bibnamefont {Javaloyes}},
  \ and\ \bibinfo {author} {\bibfnamefont {S.~V.}\ \bibnamefont {Gurevich}},\
  }\href {\doibase 10.1063/5.0002015} {\bibfield  {journal} {\bibinfo
  {journal} {Chaos}\ }\textbf {\bibinfo {volume} {30}},\ \bibinfo {pages}
  {063137} (\bibinfo {year} {2020})},\ \Eprint
  {http://arxiv.org/abs/https://doi.org/10.1063/5.0002015}
  {https://doi.org/10.1063/5.0002015} \BibitemShut {NoStop}%
\bibitem [{\citenamefont {Javaloyes}\ \emph {et~al.}(2015)\citenamefont
  {Javaloyes}, \citenamefont {Ackemann},\ and\ \citenamefont
  {Hurtado}}]{JAH-PRL-15}%
  \BibitemOpen
  \bibfield  {author} {\bibinfo {author} {\bibfnamefont {J.}~\bibnamefont
  {Javaloyes}}, \bibinfo {author} {\bibfnamefont {T.}~\bibnamefont {Ackemann}},
  \ and\ \bibinfo {author} {\bibfnamefont {A.}~\bibnamefont {Hurtado}},\ }\href
  {\doibase 10.1103/PhysRevLett.115.203901} {\bibfield  {journal} {\bibinfo
  {journal} {Phys. Rev. Lett.}\ }\textbf {\bibinfo {volume} {115}},\ \bibinfo
  {pages} {203901} (\bibinfo {year} {2015})}\BibitemShut {NoStop}%
\bibitem [{\citenamefont {Camelin}\ \emph {et~al.}(2016)\citenamefont
  {Camelin}, \citenamefont {Javaloyes}, \citenamefont {Marconi},\ and\
  \citenamefont {Giudici}}]{CJM-PRA-16}%
  \BibitemOpen
  \bibfield  {author} {\bibinfo {author} {\bibfnamefont {P.}~\bibnamefont
  {Camelin}}, \bibinfo {author} {\bibfnamefont {J.}~\bibnamefont {Javaloyes}},
  \bibinfo {author} {\bibfnamefont {M.}~\bibnamefont {Marconi}}, \ and\
  \bibinfo {author} {\bibfnamefont {M.}~\bibnamefont {Giudici}},\ }\href
  {\doibase 10.1103/PhysRevA.94.063854} {\bibfield  {journal} {\bibinfo
  {journal} {Phys. Rev. A}\ }\textbf {\bibinfo {volume} {94}},\ \bibinfo
  {pages} {063854} (\bibinfo {year} {2016})}\BibitemShut {NoStop}%
\end{thebibliography}
%

\end{document}